\documentclass[10pt,superscriptaddress,twocolumn,amsmath,amssymb,aps,prl,showpacs]{revtex4}
\usepackage{mathrsfs}
\usepackage{graphicx}
\usepackage{dcolumn}
\usepackage{bm}
\usepackage{amssymb}
\usepackage{amsmath}
\usepackage{paralist}
\usepackage[colorlinks,linkcolor=red,anchorcolor=blue,citecolor=blue]{hyperref}

\def\be{\begin{equation}}
\def\ee{\end{equation}}
\def\bea{\begin{eqnarray}}
\def\eea{\end{eqnarray}}

\newcommand{\dr}[1]{\textcolor{blue}{#1}}

\begin{document}

\title{Emergent ballistic transport of Bose-Fermi mixtures in one dimension }

\author{Sheng Wang}
\affiliation{State Key Laboratory of Magnetic Resonance and Atomic and Molecular Physics,
Wuhan Institute of Physics and Mathematics, Innovation Academy for Precision Measurement Science and Technology,  Chinese Academy of Sciences, Wuhan 430071, China}
\affiliation{University of Chinese Academy of Sciences, Beijing 100049, China}

\author{Xiangguo Yin}
\affiliation{Institute of Theoretical Physics, Shanxi University, Taiyuan 030006, China}

\author{Yang-Yang Chen}
\affiliation{Shenzhen Institute for Quantum Science and Engineering, and Department of Physics, Southern University of Science and Technology, Shenzhen 518055, China}
\affiliation{State Key Laboratory of Magnetic Resonance and Atomic and Molecular Physics,
Wuhan Institute of Physics and Mathematics, Innovation Academy for Precision Measurement Science and Technology, Chinese Academy of Sciences, Wuhan 430071, China}

\author{Yunbo Zhang}
   \email[]{ybzhang@sxu.edu.cn}
\affiliation{Institute of Theoretical Physics, Shanxi University, Taiyuan 030006, China}

\author{Xi-Wen Guan}
\email[]{xwe105@wipm.ac.cn}
\affiliation{State Key Laboratory of Magnetic Resonance and Atomic and Molecular Physics,
Wuhan Institute of Physics and Mathematics, Innovation Academy for Precision Measurement Science and Technology, Chinese Academy of Sciences, Wuhan 430071, China}
\affiliation{Department of Theoretical Physics, Research School of Physics and Engineering,
Australian National University, Canberra ACT 0200, Australia}

\date{\today}


\begin{abstract}
The degenerate Bose-Fermi (BF) mixtures in one dimension  present a novel realization
of two decoupled Luttinger liquids with bosonic and fermionic degrees of freedom at low temperatures.
However, the transport properties of such  decoupled Luttinger liquids of charges   have not yet been  studied. 
Here we apply generalized hydrodynamics to study the transport properties of  one-dimensional (1D) BF mixtures  with  delta-function interactions.
The initial state is set up as the
semi-infinite halves of two 1D BF mixtures with different temperatures, 
joined together at the time $t=0$ and the junction point $x=0$.
Using the Bethe ansatz solution, we  first rigorously prove the existence of conserved charges for 
both the  bosonic and fermionic degrees of freedom, preserving the Euler-type 
continuity  equations.
We then analytically obtain the distributions of the densities and currents of the local conserved quantities which solely depend
on  the ratio  $\xi=x/t$.
The left and right moving  quasiparticle 
excitations of the two halves form multiple segmented 
light-core hydrodynamics that display 
ballistic transport of the conserved charge densities and currents 
in different  degrees of freedom.
Our analytical  results provide a deep understanding of the quantum  transport  of  multi-component Luttinger liquids in quantum systems  with both bosonic and fermionic   statistics.

\end{abstract}
\maketitle


The nonequilibrium dynamics of one dimensional (1D) integrable systems gives much insight into 
understanding the transport of many-body systems \cite{kinoshita2006quantum,rigol2007relaxation,Iyer:2012}.
Recently, a generalized hydrodynamics (GH) approach \cite{castro2016emergent,bertini2016transport,bertini2018universal}  was introduced to  study  the evolution of macroscopic 
systems in 1D.
For integrable systems,  there are infinitely many  conserved quantities in the thermodynamic limit that essentially give  the generalized Gibbs ensemble (GGE) \cite{rigol2007relaxation,castro2016emergent,ilievski2015complete,pozsgay2014correlations} to evaluate  local observables in equilibrium.
With the help of  local equilibrium GGE,  density distributions of all conserved quantities and  their  continuity equations determine   the evolution 
of the systems.
For the two semi-infinite halves of a homogeneous 1D  system suddenly joined
at the origin, the steady state may admit nonzero stationary charge  currents inside the transient region, indicating the  ballistic transport of integrable models \cite{bernard2016conformal,castro2016emergent,doyon2017large,bertini2016transport,mestyan2019spin,bertini2018universal,bulchandani2018bethe,fagotti2017higher,bulchandani2017solvable,doyon2018soliton,ilievski2017ballistic,mazza2018energy,collura2018analytic,bastianello2018generalized,bertini2018low}.
Very recently, this  GH method  has been adapted  to study correlation functions\cite{doyon2018exact,piroli2017transport}, entanglement\cite{alba2018entanglement,alba2019entanglement,bertini2018entanglement}, Drude weight\dr{s}\cite{doyon2017drude,urichuk2019spin} and the evolution of 1D  systems confined by the  external field \cite{doyon2017note,caux2017hydrodynamics,schemmer2019generalized}.
Beyond  Euler-type hydrodynamics, diffusion  has also been analyzed with
the GH method \cite{de2018hydrodynamic,gopalakrishnan2018hydrodynamics,gopalakrishnan2019kinetic,de2018diffusion,gopalakrishnan2019anomalous,panfil2019linearized}.

On the other hand, at low temperatures, universal phenomena emerge in  quantum many-body systems.
 In 2D and 3D,  the low energy physics  of interacting fermions can be described by Landau's Fermi liquid theory,  
whereas the low energy physics  of 1D systems is elegantly  described  by the theory of Luttinger liquids \cite{Haldane:1981,Haldane:1981JPC,giamarchi2003quantum}, i.e.\ collective motions of bosons.
 %
%
In this scenario, the low energy physics of  
1D systems with  internal spin  degrees of freedom is universally described by  the  spin charge separation theory of two Luttinger liquids.
In contrast to the spin charge separated Luttinger liquids in an interacting Fermi gas \cite{Lee:2012,giamarchi2003quantum,Schulz:1991}, 
1D degenerate Bose-Fermi (BF) mixtures   display two decoupled Luttinger liquids with both  bosonic and fermionic degrees of freedom at low temperatures \cite{Lai-PhysRevA.3.393,Batchelor-PhysRevA.72.061603,Frahm-PhysRevA.72.061604,imambekov2006exactly,imambekov2006applications}.
In the former, the effective antiferromagnetic spin-spin interaction  leads to a
Luttinger liquid of spinons, whereas in the latter the Luttinger liquids  uniquely emerge in interacting particles with different quantum statistics.
It turns out that the Luttinger theory can be applied to study 
transport problems in 1D \cite{bernard2016conformal,bertini2018universal,mestyan2019spin,Nozawa:2019}.
%
%
%
%

In this paper we report on the transport properties of the 1D BF mixtures with delta-function interactions.
The initial state is taken to be
two semi-infinite  1D tubes of the BF mixture with different temperatures suddenly  joined together.
Building on the Bethe ansatz (BA)  solution,  we  rigorously prove the existence of  conserved charges for
both the  bosonic and fermionic degrees of freedom.
We  analytically obtain the distributions of the densities and currents of the local conserved quantities for the 
bosons and fermions as a function of the ratio
$\xi=x/t$. Here $x$ is the distance from the junction and $t$ is the time.
%
 %
 The quasiparticle excitations  in the two halves elegantly determine the   segmented  multiple  light-core hydrodynamics that displays the ballistic transport
 of the densities and currents  of  the conserved  charges.
The nonlinear aspect of spectrum gives rise to different  broadening features of the  light cones in distributions of densities  and currents of bosons and fermions.
 The  subtle segmented charge-charge separation emerges in  the steady state of the system 
 within the light cones of the two degrees of freedom.

 %
 %

%
%

{\bf The model  and conserved charges.}
A Bose-Fermi mixture system is described by the  Hamiltonian \cite{imambekov2006exactly,imambekov2006applications}
\begin{eqnarray}
H&=&\int_{0}^{L}dx\left( \frac{\hbar^2}{2m_b}\partial_x \psi_b^{\dagger}\partial_x\psi_b+\frac{\hbar^2}{2m_f}\partial_x \psi_f^{\dagger}\partial_x\psi_f\right.\nonumber\\
&&\left.+\frac{g_{bb}}{2}\psi_b^{\dagger}\psi_b^{\dagger}\psi_b\psi_b+g_{bf}\psi_b^{\dagger}\psi_f^{\dagger}\psi_f\psi_b\right),\label{Ham}
\label{Hamiltonian}
\end{eqnarray}
where $\psi_b$ and $\psi_f$ are boson and fermion field operators and
$m_b$ and $m_f$ are the masses of the bosons and fermions, respectively.  We denote by
$g_{bb}$ and $g_{bf}$   the boson-boson and boson-fermion interaction strengths, noting that their fermion-fermion counterpart is zero
due to the Pauli exclusion statistics of spinless fermions.  When $m_b=m_f=m$ and  $g_{bb}=g_{bf}=g$, the system is integrable  \cite{Lai-PhysRevA.3.393,Batchelor-PhysRevA.72.061603,Frahm-PhysRevA.72.061604,imambekov2006exactly,imambekov2006applications}.

We consider 
a system with $N$ total particles of which
$M$ are bosons and  $N-M$ are spinless fermions.
The model  can be solved using a nested Bethe Ansatz in the framework of the quantum inverse scattering method, see the supplementary material \cite{SM}.
The spectrum   $E=\sum_{i=1}^Nk_i^2$  of the Hamiltonian (\ref{Ham}) is determined  by the  quasi-momenta $\left\{k_i\right\}$ with $i=1,\ldots, N$  and rapidities $\left\{\Lambda_\alpha \right\}$ with $\alpha =1, \ldots, M$ subject to  the Bethe Ansatz equations (BAE)
$ \prod_{\alpha=1}^{M}\theta \left(k_i-\Lambda_\alpha\right)=e^{ik_iL}$, and $ \prod_{j=1}^{N}\theta \left(k_j-\Lambda_\alpha\right)=1$,
where  $\theta(x) =\left( x+\frac{\mathrm{i} c}{2}\right) /\left(x-\frac{\mathrm{i} c}{2}\right)$.
The  solutions of the BAE give all the eigenstates.

We  first prove  the existence of two independent sets of conserved quantities of the BF mixture   based on the polynomial form of its quantum  transfer matrix, see \cite{SM}.
The eigenvalues of the transfer matrix are
given by the nested Bethe Ansatz
\begin{eqnarray}
t(u)=\prod_{\alpha=1}^{M}\left( \frac{u-\Lambda_\alpha+ic}{u-\Lambda_\alpha}\right) \left(1-\frac{\prod_{j=1}^{N}(u-k_j)}{\prod_{j=1}^{N}{(u-k_j-ic)}} \right),
\end{eqnarray}
exhibiting them as rational functions of the spectral parameter $u$.
 All the
 coefficients of $t(u)$, when expanded as a polynomial in $u$ and $u^{-1}$, are 
 conserved quantities of the system.
These coefficients may
be expressed as a product of 
two parts. One
involves only the $\left\{k_i\right\}$, whereas the other
involves only the $\left\{\Lambda_\alpha \right\}$.
 The integrability suggests  \cite{imambekov2006exactly,imambekov2006applications} that there are $N$ conserved quantities of the first type. 
  Thus  the remaining
 $M$ coefficients, i.e.\ those  involving only the $\left\{\Lambda_\alpha \right\}$, are the independent conserved quantities in the bosonic degree of freedom.

In the thermodynamic limit,  the   BAE  \cite{yin2009yang} can be  written in terms of the distributions of particles $\rho(k), \sigma (\Lambda) $  and holes $\rho_h(k), \sigma_h(\Lambda) $  in the total charge and  bosonic sectors
\begin{eqnarray}
&& \rho_t(k) =\frac{1}{2\pi}+a(k) \ast \sigma (\Lambda),\nonumber\\
&&  \sigma_t(\Lambda)=a(\Lambda) \ast \rho(k), \label{BAE}
\end{eqnarray}
where $a(x)=\frac{1}{2\pi}\frac{4c}{c^2+4x^2}$  and $\ast$  denotes the convolution $a(x)\ast f(y)=\int_{-\infty}^{\infty} a(x-y)f(y)dy$.
Following the Yang-Yang thermodynamic approach \cite{Yang-Yang1969},
we obtain the thermodynamic Bethe Ansatz equations (TBAE) for equilibrium states  at finite temperatures $T$ \cite{yin2009yang,yin2012quantum},
\begin{eqnarray}
\varepsilon(k)&=&k^2-\mu_f- a(k)\ast f_{\varphi} (\Lambda),\nonumber\\
\varphi(\Lambda)&=&\mu_f-\mu_b-a(\Lambda)\ast f_\varepsilon (k), \label{TBAE}
\end{eqnarray}
where we denoted by $\varepsilon(k)$ and $\varphi(\Lambda)$ 
the energies of the excitations, $f_{\varepsilon,\varphi}(x)=T\ln \left(1+e^{-\frac{\varepsilon,\varphi(x)}{T}} \right)$ and  $\mu_f$ and $\mu_b$ are the chemical potentials of the fermions and bosons.
The BAE (\ref{BAE}) and the TBAE (\ref{TBAE}) provide a convenient setting for determining  the distributions of densities and  currents of all conserved charges in the two degrees of freedom.

{\bf Generalized Hydrodynamics.}
%
 %
 We apply the GH  \cite{castro2016emergent,bertini2016transport,bertini2018universal} to calculate the distributions of densities and  currents of all conserved charges of the model.
 To this end, we denote the two sets of single-particle conserved quantities by
 $h_\rho^i(k)$ and $h_\sigma^i(\Lambda)$. For example:
$h_\rho^1(k)=1$; $h_\rho^2(k)=p(k)=k$; $h_\rho^3(k)=e(k)=k^2$; $h_\sigma^1(\Lambda)=1$, etc.
Analogously, under the local equilibrium approximation,  a non-equilibrium state of an integrable system
can be identified by the distributions of all the conserved quantities  $q^1(x),q^2(x),...,q^N(x),...,x\in [0,L]$, where $q^i=\langle\hat{q}^i\rangle$ ,
$\hat{Q}^i=\int dx \hat{q}^i(x)$ and $\hat{Q}^i$ is the operator of the conserved quantity.
Then, we have 
expressions for the densities of all the conserved quantities
$q^i_{b} (x,t)=\int dk\, b (k,x,t)h_b^i(k), \quad b=\rho, \sigma$.
Following the GH approach  \cite{de2018diffusion},  the expectation value of the currents is
given by
$ j^i_b(x,t)=\int dkb(k,x,t)\,v_b  (k,x,t)h_b^i(k)$
 for a homogeneous stationary state.
Here $v_b$ (with $b=\rho,\sigma$) 
is
the sound velocity
of the excitations in the total charge and the bosonic degrees of freedom with
$v_\rho(k)=\frac{\partial \varepsilon(k)}{\partial p_\rho^{dr}(k)}=\frac{ \varepsilon'(k)}{p_\rho'^{dr}(k)}$ 
and
$v_\sigma(\Lambda)=\frac{\partial \varphi(\Lambda)}{\partial p_\sigma^{dr}(\Lambda)}=\frac{\varphi'(\Lambda)}{p_\sigma'^{dr}(\Lambda)}$.
Here  $p_\rho^{dr}(k)$ and  $p_\sigma^{dr}(\Lambda)$ denote 
the dressed momenta of the excitations.

Building on  the   densities and currents of all conserved quantities in terms of Bethe ansatz  densities via the above definitions of $q^i_{b} (x,t)$ and $j^i_b(x,t)$, and using the BAE (\ref{BAE}) and 
TBAE (\ref{TBAE}),  we  can prove the continuity equations for the total charge and the bosonic degrees of freedom \cite{SM}
\begin{eqnarray}
&&\partial_tn_\rho (k,x,t)+v_\rho (k,x,t)\partial_xn_\rho (k,x,t)=0, \nonumber\\
&&\partial_tn_\sigma(\Lambda,x,t)+v_\sigma(\Lambda,x,t)\partial_xn_\sigma(\Lambda,x,t)=0, \label{SICE}
\end{eqnarray}
where we defined the occupation numbers
$n_\rho(k)=\frac{\rho(k)}{\rho(k)+\rho_h(k)}$ 
and $\quad n_\sigma(\Lambda)=\frac{\sigma(\Lambda)}{\sigma(\Lambda)+\sigma_h(\Lambda)}$.

\begin{widetext}

\begin{figure}[t]
	\begin{center}
		\includegraphics[scale=0.35]{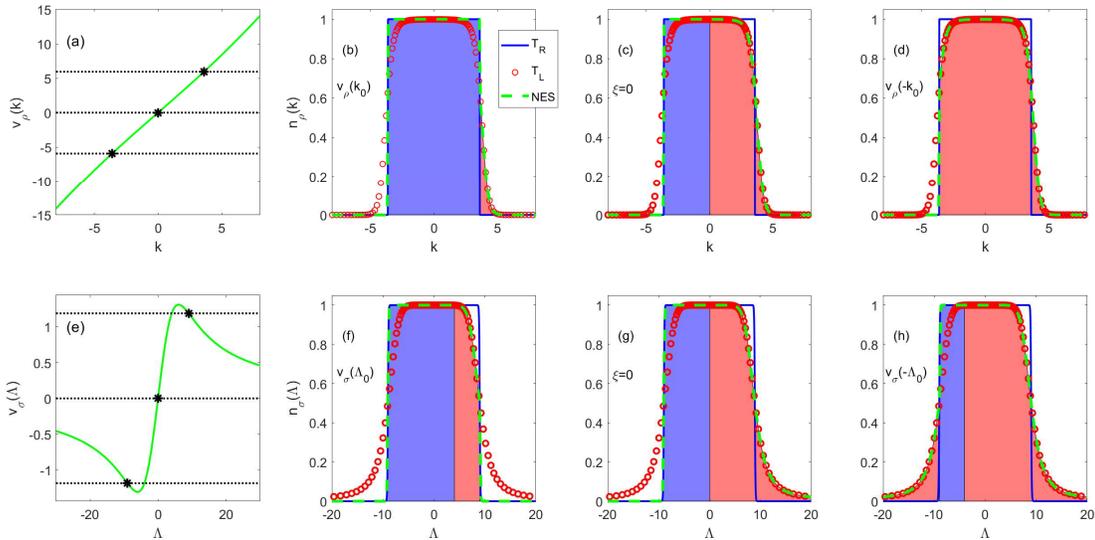}
\end{center}
\caption{ (a) and (e) show the velocities of the charge and the bosonic excitations, respectively. Here the asterisks correspond to the zero and the Fermi quasimomenta  $k=0, \, \pm k_0$ and $\Lambda =0,\, \pm\Lambda_0$, respectively.
 (b-d)  The occupation number $n_\rho$ of the  total charge 
 is  symmetric  for the equilibrium state in  the two halves:  the blue  solid  line indicates  the right half  at $T_R=0.01$ and the red  circles indicate
 the left half  at $T_L=2$. The occupation number of the total charges is asymmetric   for the  non-equilibrium state (NES).    The green  dashed lines  show the occupation numbers  determined by  Eq.~(\ref{final_solutions}):  (b) for   $\xi=v_{\rho}\left(k_{0}\right)$, (c) for  $\xi=0$ and (d) for $\xi=v_{\rho}\left(-k_{0}\right)$, respectively.
(f-h) The occupation number of the bosons shows a  symmetric feature  for the equilibrium states  in  the two halves:  again a blue
solid line for the right half at $T_R=0.01$ and red circles for the left half
at $T_L=0.2$.  The green  dashed lines  show the occupation numbers  determined by the NES  Eq.~(\ref{final_solutions}):
(f) for $\xi=v_{\sigma}\left(\Lambda_{0}\right)$,  (g) for $\xi=0$ and (h) for $\xi=v_{\sigma}\left(-\Lambda_{0}\right)$, respectively.
The filling areas for the NES (green dashed 
lines)  in (b) and (h)  are  larger than the areas for the  equilibrium states in the right and left halves, respectively,   indicating  peaks in the density profiles.
On the other hand,  the filling areas  for the NES (green dashed
lines)  in (d) and (f) are smaller than the  areas for the  equilibrium states in the left and right halves, respectively, indicating
valleys in the density profiles. (c) and (g) show the occupation numbers for the steady state, indicating  the average in the density profiles.  $\mu_b=11$, $\mu_f=12$, $c=10$ for all diagrams.
 }
\label{occupation}
\end{figure}

\end{widetext}

\begin{figure}[t]
	\begin{center}
		\includegraphics[scale=0.6]{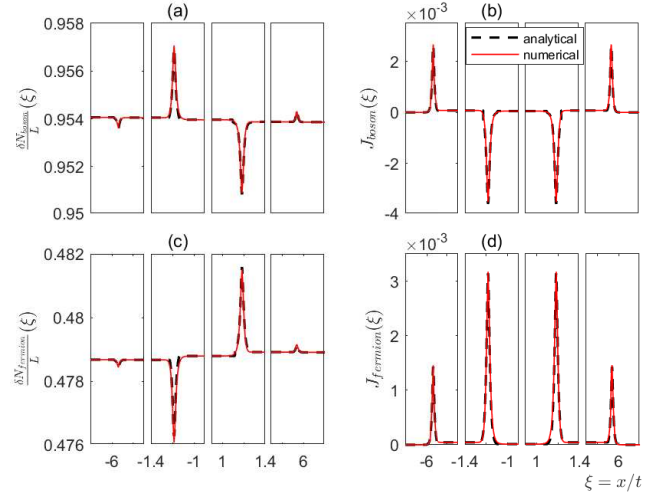}
	\end{center}
	\caption{Upper panel (a), (b) (lower panel (c), (d)) respectively show the distributions  of the density and current of bosons (fermions) at low temperatures. The dashed lines show  the results from the first order of the analytical low-temperature expansion (\ref{density}),  (\ref{eq:particle_current}) and are in excellent agreement with  the  numerical results (solid lines), see \cite{SM}.  Good agreement is also 
	seen for  the round peaks  (\ref{eq:density-peak}) and  (\ref{eq:particle_current2}) near  the points $\xi=\pm v_{c,b}$. The back flows were 
	observed at the points $\xi=\pm v_{b}$. These results were obtained with 
	$T_L=0.04$, $T_R=0.01$, $\mu_b=11$, $\mu_f=12$ and 
	$c=10$. }
	\label{density-current-profiles}
\end{figure}

We choose a special initial state, i.e.\  two semi-infinite 1D systems in equilibrium 
with $T_L>T_R$ that are suddenly joined  together, and then we let the whole system evolve under the Hamiltonian $H$ (\ref{Ham}).
Obviously, the initial state is not in  equilibrium 
and the conserved charges transport from left to right.  In view of the fact that 
the relaxation  times in the fluid cells are much less than variational scales of the physical properties \cite{castro2016emergent,bertini2016transport},  the  system has  a special  scaling invariant property, i.e.\ under the scaling transformation  $(x,t)=(\alpha x,\alpha t)$,  the GH  equations (\ref{SICE}) remain
invariant. We then have the solution
\begin{eqnarray}
&&(v_\rho(k,\xi)-\xi)\partial_\xi n_\rho(k,\xi)=0\nonumber\\
&&(v_\sigma(\Lambda,\xi)-\xi)\partial_\xi n_\sigma(\Lambda,\xi)=0  \label{SICE-2}
\end{eqnarray}
with  $\xi=x/t$.
From the initial state, we have the boundary conditions
$\lim_{\xi\to\pm \infty}n_\rho(k,\xi)=n^{R,L}_\rho(k)$, and
$\lim_{\xi\to\pm \infty}n_\sigma(\Lambda,\xi)=n^{R,L}_\sigma(\Lambda)$,
which give the flowing  solutions of
Eq.~(\ref{SICE-2})
\begin{eqnarray}
n_\rho(k,\xi)=n^L_\rho(k)H(v_\rho(k,\xi)-\xi)+n^R_\rho(k)H(\xi-v_\rho(k,\xi)),\nonumber \\
n_\sigma(\Lambda,\xi)=n^L_\sigma(\Lambda)H(v_\sigma(\Lambda,\xi)-\xi)+n^R_\sigma(\Lambda)H(\xi-v_\sigma(\Lambda,\xi)), \label{final_solutions}
\end{eqnarray}
where $H(x)$ denotes the Heaviside step function, $H(x)=1$ for $x\ge 0$ and zero otherwise.
In Fig.~\ref{occupation}, we show occupation numbers in the charge and bosonic sectors for different values of $\xi$.
In (b)-(d) and (f)-(h),  the filling areas show the occupation numbers of the total charges and bosons  in both  equilibrium states and NES, determining  the density profiles in transport, also see Fig. ~\ref{density-current-profiles}.
Fig.~\ref{occupation} (a) shows that the effective velocity of excitations in the charge sector  monotonously  increases as the
quasi-momentum increases,  indicating positive effective mass in the nonlinear dispersion.
However,  in Fig.~\ref{occupation} (e), the effective velocity of the bosons  is observed to be negative  for 
large $|\Lambda|$, indicating negative effective mass in the nonlinear dispersion.
Such a subtle negative mass  leads to a back flow in the boson density current
at the light cone $\xi=v_\sigma(\Lambda_0)$, see Fig.~\ref{density-current-profiles} (b).

\begin{figure}[t]
	\begin{center}
		\includegraphics[scale=0.53]{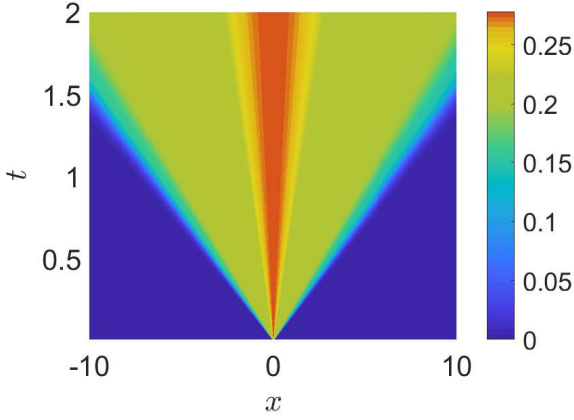}
	\end{center}
	\caption{The light cone diagram of the of the energy current of   the 1D BF mixture after 
	suddenly joining the wo semi-infinite halves  at different low  temperatures.
	Here we set  $T_L=1$, $T_R=0.5$, $\mu_b=11$,  $\mu_f=12$  and $c=10$.  The steady state regions (constant value)  and transition regions (peaks or valleys) show the distinguished features of the  separated linear and non-linear Luttinger  liquids, respectively, see the text. }
	\label{light-cone}
\end{figure}

{\bf Emergent ballistic transport.}  Calculating the  GH   equations (\ref{final_solutions}) is always 
analytically difficult.
At low temperatures, the distributions of the densities and currents of all conserved charges, i.e.\ both the charge and  bosonic degrees of freedom,  can be expressed as  polynomials in 
the temperature, see \cite{SM}.
   The  excitation energy density distribution relative to  
   the ground state is given by
$\delta E/L=\sum_{i=c,b}\frac{\pi}{6v_{i}}W_{i}$,
where $v_{c}=v_{\rho}\left(k_{0}\right)$ and
$v_{b}=v_{\sigma} \left(\Lambda_{0}\right)$ are the Fermi  velocities at $T=0$.
The functions $W_{i}$  are given by
$ W_{i}=T_{R}^{2}H\left(\xi-v_{i}\right)+T_{L}^{2}H\left(-v_{i}-\xi\right)+\frac{1}{2}\left(T_{L}^{2}+T_{R}^{2}\right)H\left(v_{i}-\left|\xi\right|\right)$
 with  $i=c,b$.  We thus  immediately obtain the excitation  energy density and current for the steady state
\begin{eqnarray}
\frac{\delta E_{\textrm{steady}}}{L}&=&\frac{1}{2}\left(T_{L}^{2}+T_{R}^{2}\right)\sum_{i=c,b}\frac{\pi}{6v_{i}}, \,\, {\rm for}\,\, \xi=0 \\
j^{E}&=&\frac{\pi^{2}}{6}\left(T_{L}^{2}-T_{R}^{2}\right)\sum_{i=c,b}H\left(v_{i}-\left|\xi\right|\right).\label{eq:energy_current}
\end{eqnarray}
These  show that the multiple light cone evolutions of  the energy density and current  are  determined by the ballistic transport of the quasiparticles with  local velocities $v_\rho(k)$ and $v_\sigma(\Lambda)$ in the two sectors.
Solving the GH equations (\ref{final_solutions}), the
BAE (\ref{BAE}) and the TBAE (\ref{TBAE}) by iteration, we may obtain the distributions of  energy, densities and currents of the conserved  charges.
Fig.~\ref{light-cone} elegantly presents the
ballistic transport of the energy current within  different light cones.
In this figure,  we clearly observe the  crossover of  broadening  in the vicinity
of the light cone, confirming  the prediction from 
Luttinger liquid theory  \cite{bertini2018universal}.

The change in the density 
of total charges is 
obtained in a unified  form
\begin{equation}
 \delta N/L=\sum_{i=c,b}F_{i} W_{i}\label{density}
 \end{equation}
for the regions $\left|\xi-r v_{i}\right|>T_{L\left(R\right)}/m_{i}^{*}v_{i}$ with $i=c,b$ and $r=\pm1 $.
In the above equations
$
F_{i}=\frac{\pi}{6K_i v_i }\left[f_i' \left(\lambda_{0}\right)-\left(\frac{v_{i}'}{v_{i}}+A\left(\lambda_{0}\right)\right)f_{i}\left(\lambda_{0}\right)\right].
$
Here $K_c=\varepsilon'_0(k_0)$, $K_b=\varphi_0'(\Lambda_0)$,  $v_{c}'=\left(v_{\rho}\right)'\left(k_{0}\right)$ and
$v_{b}'=\left(v_{\sigma}\right)'\left(\Lambda_{0}\right)$, while
$f'_i(\lambda_0)=\left[ \partial f_i(\lambda ) /\partial \lambda \right]_{\lambda =\lambda_0}$
and $f_i(\lambda )$ and $A_i(\lambda )$ satisfy the following equations
\begin{eqnarray}
f_i(\lambda) &=&f_i^{\rm bare} +\int_{-\lambda_0}^{\lambda_0}a (\lambda -\Lambda) f_{\bar{i}}(\Lambda) d\Lambda,\label{function-A-f}\\
\frac{K_{\bar{i} } }{K_i}A_i (\lambda ) &=&\int_{-\lambda_0}^{\lambda_0} a(\lambda-\Lambda) A_{\bar{i}}(\Lambda )d\Lambda -\sum_{\ell =\pm1}a(\lambda +\ell \lambda_0)  \nonumber
\end{eqnarray}
with $f^{\rm bare}_{c(b)}=1(0)$, and $\bar{c}=b,\,\bar{b}=c$, respectively.

In contrast,  for  the crossover region $\left|\xi-rv_{i}\right|<T_{L\left(R\right)}/m_{i}^{*}v_{i}$ with $i=c,b$ and $r=\pm1 $, the nonlinear dispersion effect  gives rise to  the peak/valley densities of charges
\begin{equation}
\frac{\delta N_i^{r }}{L}=\textrm{sgn}\left(rm_{i}^{*}\right)\frac{T_Lf_i(\lambda_0)}{2\pi v_i} D_{\frac{T_{R}}{T_{L}}}\left[\frac{m_{i}^{*}v_{i}}{T_{L}}\left(\xi-rv_{i}\right)\right]\label{eq:density-peak}
\end{equation}
with   $D_{\eta}\left(z\right)\equiv\ln\left(1+e^{z}\right)-\eta\ln\left(1+e^{z/\eta}\right)$. Here  the effective mass is  defined by
$m_{i}^{*}=K_i/(v_i'v_i)$.
The Eq.~(\ref{eq:density-peak}) shows a universal broadening of the light cones in the low temperature hydrodynamics of a 1D interacting BF
mixture in both the charge and bosonic degrees of freedom.

We can moreover derive the particle  current for the total charges in the transient  region $\left|\xi-r v_{i}\right|>T_{L\left(R\right)}/m_{i}^{*}v_{i}$ with  $i=c,b$ and $r=\pm1 $,
\begin{equation}
j=\left(T_{L}^{2}-T_{R}^{2}\right)\sum_{i=c,b}G_{i}H\left(v_{i}-\left|\xi\right|\right)\label{eq:particle_current}
\end{equation}
with
$G_{i}=\frac{\pi}{12K_i}\left[f_i'(\lambda_0)-A_i(\lambda_0) f_i (\lambda_0)  \right]$.
Similarly, in the vicinity 
of the light cones  $\left|\xi-rv_{i}\right|<T_{L\left(R\right)}/m_{i}^{*}v_{i}$ with  $i=c,b$ and $r=\pm1 $,  the peaks (or valleys ) of the particle current are given by
\begin{equation}
j_i^{r }=\textrm{sgn}\left(m_{i}^{*}\right)\frac{T_Lf_i(\lambda_0)}{2\pi} D_{\frac{T_{R}}{T_{L}}}\left[\frac{m_{i}^{*}v_{i}}{T_{L}}\left(\xi-rv_{i}\right)\right].
\label{eq:particle_current2}
\end{equation}
From the  results (\ref{density}), (\ref{eq:density-peak}), (\ref{eq:particle_current}) and (\ref{eq:particle_current2}),  we can directly obtain the distributions of  the density and current of bosons and fermions which show excellent agreement 
with numerical calculations, see Fig.~\ref{density-current-profiles}.  Here we noted 
that the total particle number is conserved and used 
the function $f^{\rm bare}_{c(b)}=0(1)$ in  Eq.~(\ref{function-A-f})   for the bosonic degree of freedom.

In summary,
we have presented universal emergent  transport properties of the 1D BF mixture, in which two decoupled Luttinger liquids are formed in the bosonic and fermionic degrees of freedom, respectively.
We have analytically obtained  the distributions of the densities and currents of the local conserved quantities that show the ballistic transport 
of the quasiparticles in both charge degrees. At low temperatures, our  analytical expressions for the
transport properties provide a microscopic origin of 
ballistic transport for integrable systems with both bosonic and fermionic fields. 
This reveals the role of  interaction and  quantum statistical effects in  quantum transport.

\section*{Acknowledgement}
The authors thank David Ridout  for going through the manuscript and also  thank Yuzhu Jiang, M\'{a}rton Mesty\'{a}n, Giuseppe Mussardo,   Han Pu and Randy Hulet  for helpful discussions.
This work is supported by   the National Key R\&D Program of China  No.\ 2017YFA0304500,  the key NSFC grant No.\ 11534014 No.\ 11874393 and No.\ 11674201.
XWG acknowledges Rice University for
supporting his visit.


\clearpage\newpage
\setcounter{figure}{0}
\setcounter{table}{0}
\setcounter{equation}{0}
\def\thefigure{S\arabic{figure}}
\def\thetable{S\arabic{table}}
\def\theequation{S\arabic{equation}}
\setcounter{page}{1}
\pagestyle{plain}

\begin{widetext}

\section*{Emergent ballistic transport of Bose-Fermi mixtures in one dimension}

\begin{center}
\noindent{ Sheng Wang, Xiangguo Yin, Yang-Yang Chen, Yunbo Zhang and Xi-Wen Gaun}
\end{center}

\section{I. Bethe ansatz and quantum inverse scattering method}
Here, we give the exact solutions of Bose-Fermi mixture by nested Bethe Ansatz\cite{yang1967some,lieb1968absence,faddeev1996algebraic}

Wavefunction of the system is supposed to be symmetric with respect to insices $i={1,...,M}$(bosons) and antisymmetric with respect to $i={M+1,...,N}$(fermions). 

We can assume the coordinate Bethe wavefunction have the following form: for $0<x_{Q_1}<x_{Q_2}<...<x_{Q_N}<L$
\begin{equation}
\Psi_\sigma(x)=\sum_{P}A_\sigma(P,Q)e^{i\sum_{i}k_{P_i}x_{Q_i}},
\end{equation}
where $\sigma=(\sigma_1,\sigma_2,...,\sigma_N)$, with $\sigma_j$ denoting the $SU(1|1)$ component of the $j$th particles; $P,Q$ are arbitrary permutations from $S_N$.

From Schrodinger equation, we have
\begin{equation}\label{}
(\frac{\partial}{\partial x_j}-\frac{\partial}{\partial x_k})\Psi_{x_j=x_k^+}-(\frac{\partial}{\partial x_j}-\frac{\partial}{\partial x_k})\Psi_{x_j=x_k^-}=2c\Psi_{x_j=x_k}
\end{equation}
and from the continuity of wavefunction, we have 
\begin{equation}
\Psi_{x_j-x_k=0^+}=\Psi_{x_j-x_k=0^-}.
\end{equation}

Suppose $Q$ and $Q'$ are two permutations, such that 
\begin{equation}
Q=(...Q_aQ_b...), Q'=(...Q_bQ_a...), b=a+1.
\end{equation}
Similarly, $P$ and $P'$ are two permutations,
\begin{equation}
P=(...P_aP_b...), P'=(...P_bP_a...).
\end{equation}

Then, for the different regions $Q$ and $Q'$, we have the Bethe wavefunctions as the following forms,
\begin{align}
&Q: \Psi_\sigma(x)=\sum\left\lbrace A_\sigma(P,Q)e^{...+ik_{P_a}x_{Q_a}+ik_{P_b}x_{Q_b}+...}+A_\sigma(P',Q)e^{...+ik_{P_b}x_{Q_a}+ik_{P_a}x_{Q_b}+...}\right\rbrace, \\
&Q': \Psi_\sigma(x)=\sum\left\lbrace A_\sigma(P,Q')e^{...+ik_{P_a}x_{Q_b}+ik_{P_b}x_{Q_a}+...}+A_\sigma(P',Q')e^{...+ik_{P_b}x_{Q_b}+ik_{P_a}x_{Q_a}+...}\right\rbrace.
\end{align}
According to the two conditions above, we have
\begin{equation}
A_\sigma(P,Q)=\frac{(k_{P_a}-k_{P_b})P_{Q_aQ_b}+ic}{k_{P_a}-k_{P_b}-ic}A_\sigma(P',Q).
\end{equation}
where 
\begin{equation}
A_\sigma(P',Q')=P_{Q_aQ_b}A_\sigma(P',Q).
\end{equation}
For simplifying the equation, we set
\begin{equation}
Y^{ab}_{P_bP_a}=\frac{(k_{P_a}-k_{P_b})P_{Q_aQ_b}+ic}{k_{P_a}-k_{P_b}-ic}.
\end{equation}

We can also define the permutation operator $P_{\sigma_{Q_a}\sigma_{Q_b}}$ of two $\sigma_{Q_a}\sigma_{Q_b}$, such that 
\begin{align}
&P_{\sigma_{Q_a}\sigma_{Q_b}}P_{Q_aQ_b}=1, \text
{  (identical principle) }\\
A_{...\sigma_{Q_a}...\sigma_{Q_b}...}(P',Q)=P_{\sigma_{Q_a}\sigma_{Q_b}}&P_{Q_aQ_b}A_{...\sigma_{Q_a}...\sigma_{Q_b}...}(P',Q)=P_{\sigma_{Q_a}\sigma_{Q_b}}A_{...\sigma_{Q_a}...\sigma_{Q_b}...}(P',Q').
\end{align}
Then we have 
\begin{equation}
\begin{aligned}
A_\sigma(P,Q)&=Y^{ab}_{P_bP_a}A_\sigma(P',Q)=\frac{(k_{P_b}-k_{P_a})-icP_{\sigma_{Q_a}\sigma_{Q_b}}}{k_{P_b}-k_{P_a}+ic}A_\sigma(P',Q').
\end{aligned}
\end{equation}
Finally, we have the two-body scattering operator,
\begin{equation}
S_{Q_aQ_b}(k_{P_b}-k_{P_a})=\frac{(k_{P_b}-k_{P_a})-icP_{\sigma_{Q_a}\sigma_{Q_b}}}{k_{P_b}-k_{P_a}+ic}.
\end{equation}

To solve the problem of $N$ particles in a box of length $L$, we need the boundary condition, and here we apply the periodic boundary condition
\begin{equation}
\Psi(x_1,...,x_i,...,x_N)=\Psi(x_1,...,x_i+L,...,x_N)
\end{equation}
on the wavefunction with period $L$ for every $1\le i\le N$.

Without  losing  generality, we can take the position of the $i$th particle to be 0, $x_i=0$. For the wavefunction to be confined in the region $0\le x_1\le x_2\le ...\le x_N\le L$, we have
\begin{equation}
\Psi(x_i=0,x_1,...,x_N)=\Psi(x_1,...,x_N,x_i=L)
\end{equation}
Writing out both wavefunctions explicitly,
\begin{equation}
\sum_{P}A_\sigma(P_1,...P_N;Q_i,Q_1,...,Q_N)e^{i(k_{P_1}x_{Q_i}+...+k_{P_N}x_{Q_N})}=\sum_{P'}A_\sigma(P'_1,...P'_N;Q_1,...,Q_N,Q_i)e^{i(k_{P'_1}x_{Q_1}+...+k_{P'_N}x_{Q_i})}.
\end{equation}
To compare terms on  both sides of the equation with the same exponent, we have
\begin{equation}
A_\sigma(P_i,P_1,...P_N;Q_i,Q_1,...,Q_N)=e^{ik_iL}A_\sigma(P_1,...P_N,P_i;Q_1,...,Q_N,Q_i),
\end{equation}
\begin{equation}
LHS=S_{1,i}(k_1-k_i)S_{2,i}(k_2-k_i)...S_{i-1,i}(k_{i-1}-k_i)A_\sigma(P,Q),
\end{equation}
\begin{equation}
RHS=S_{i,N}(k_i-k_N)S_{i,N-1}(k_i-k_{N-1})...S_{i,i+1}(k_{i}-k_{i+1})e^{ik_iL}A_\sigma(P,Q).
\end{equation}
Then we have
\begin{equation}\label{eigenvalue_equation}
S_{i+1,i}(k_{i+1}-k_i)...S_{N,i}(k_N-k_i)S_{1,i}(k_1-k_i)...S_{i-1,i}(k_{i-1}-k_i)A_\sigma(P,Q)=e^{ik_iL}A_\sigma(P,Q).
\end{equation}

Consequently, to obtain the exact solutions of the system, we need to solve the eigenvalue equation above. Algebraic Bethe Ansatz is very successful in doing  this problem.  
Considering the local  space as the Hilbert space $h_n=C^2$, then the entire state space of the system with $N$ particles is given by $H_N=\prod_{n=1}^{N}\otimes h_n$.

Let us first introduce the operator $R_{ij}(u)$ 
\begin{equation}
R_{ij}(u)=\frac{u-icP_{ij}}{u+ic}
\end{equation}
which acts on the space $V_i\otimes V_j$. $V$is auxiliary space, $V=C^2$, and $P_{ij}$ is permutation operator.

Denoting the fermion and boson states as
\begin{equation}
|\uparrow\rangle=\left( 
\begin{array}{ccc}
1\\
0\\
\end{array}
\right) ,  
|\downarrow\rangle=\left( 
\begin{array}{ccc}
0\\
1\\
\end{array}
\right).
\end{equation}
Then, all the operators can be expressed as matrixs.

For Bose-Fermi mixture system, 
\begin{equation}
P_{ij}=\left( 
\begin{array}{cccc}
-1&0&0&0\\
0&0&1&0\\
0&1&0&0\\
0&0&0&1\\
\end{array}
\right), 
\end{equation}
\begin{equation}
R_{ij}(u)=\frac{u-icP_{ij}}{u+ic}=\left( 
\begin{array}{cccc}
1&0&0&0\\
0&\frac{u}{u+ic}&\frac{-ic}{u+ic}&0\\
0&\frac{-ic}{u+ic}&\frac{u}{u+ic}&0\\
0&0&0&\frac{u-ic}{u+ic}\\
\end{array}
\right). 
\end{equation}
Setting 
\begin{equation}
f(u)=\frac{u}{u+ic}; g(u)=\frac{-ic}{u+ic}; h(u)=\frac{u-ic}{u+ic}.
\end{equation}

The definition of the Lax operator on $j$ site involves the local quantum space $h_j$ and the auxiliary space $V$ which is $C^2$. It acts on  $h_j\otimes V$ and its explicit expression is given by
\begin{equation}
L_{j}(k_j-u)=f(k_j-u)+g(k_j-u)P_{j\tau}
=\left( 
\begin{array}{cc}
a(k_j-u)&b(k_j-u)\\
c(k_j-u)&d(k_j-u)\\
\end{array}
\right), 
\end{equation}
where $a(k_j-u),b(k_j-u),c(k_j-u),d(k_j-u)$ are operators which act in $h_j$.
\begin{equation}
L_{j}(k_j-u)|\uparrow\rangle=\left( 
\begin{array}{cc}
1&g(k_j-u)\sigma_j^-\\
0&f(k_j-u)\\
\end{array}
\right) |\uparrow\rangle.
\end{equation}
The Monodromy operator of the system is defined as
\begin{equation}
T_N(u)=L_{N}(k_N-u)...L_1(k_1-u)=\left( 
\begin{array}{cc}
A(u)&B(u)\\
C(u)&D(u)\\
\end{array}
\right).
\end{equation}

For Bose-Fermi mixture system, the Lax operator satisfies the graded Yang-Baxter ralation and the Monodromy operator satisfies the graded RTT relation
\begin{equation}
\widetilde{R}(u-v)[T'_N(u)T'_N(v)]=[T'_N(v)T'_N(u)]\widetilde{R}(u-v),
\end{equation}
where
\begin{equation}
\begin{aligned}
\widetilde{R}&=PR\\
T'_N(u)&=T_N(u)\otimes_sI\\
T'_N(v)&=I\otimes_sT_N(v).\\
\end{aligned}
\end{equation}
Here $\otimes_s$ is the graded tensor product 
\begin{equation}
(A\otimes_s B)^{ab}_{cd}=(-1)^{(p(a)+p(c))p(b)}A^a_cB^b_d,
\end{equation}
where $p(i)$ is Grassmann parities.

The graded ransfer matrix is defined as
\begin{equation}
\tau(u)=str_\tau T_N(u)=A(u)-D(u)
\end{equation}
Then we can obtain the equivalent equation to  (\ref{eigenvalue_equation}) as following,
\begin{equation}
\tau(k_j)A_\sigma(P,Q)=e^{ik_iL}A_\sigma(P,Q)
\end{equation}
So, what we need to do next is  diagonalizing $\tau(u)$.

The definition of ground state is given by
\begin{equation}
|\Omega\rangle=|\uparrow\rangle_N...|\uparrow\rangle_1
\end{equation}
\begin{equation}
T_N(u)|\Omega\rangle=\left( 
\begin{array}{cc}
1&B(u)\\
0&\prod_{j=1}^{N}f(k_j-u)\\
\end{array}
\right)|\Omega\rangle.
\end{equation}
From RTT relation, we can obtain the commutation relations of the matrix elements of the monodromy matrix
\begin{equation}
\begin{aligned}
&[B(u),B(v)]=0,\\
A(u)B(v)&=\frac{1}{f(u-v)}B(v)A(u)-\frac{g(u-v)}{f(u-v)}B(u)A(v),\\
D(u)B(v)&=\frac{h(v-u)}{f(v-u)}B(v)D(u)+\frac{g(v-u)}{f(v-u)}B(u)D(v).\\
\end{aligned}
\end{equation}

We assume that an arbitrary eigenstate of the system has the form $B(\Lambda_1)...B(\Lambda_M)|\Omega\rangle$, where $\Omega\rangle$ is the vacuum state and  there are a creation of  $M$  Bosons. Then let the transfer matrix act on the  eigenstate
\begin{equation}
\begin{aligned}
\tau(u)\prod_{j=1}^{M}&B(\Lambda_j)|\Omega\rangle=\left( \prod_{\alpha=1}^{M}\frac{1}{f(u-\Lambda_\alpha)}-\prod_{\alpha=1}^{M}\frac{h(\Lambda_\alpha-u)}{f(\Lambda_\alpha-u)}\prod_{j=1}^{N}f(k_j-u)\right) \prod_{j=1}^{M}B(\Lambda_j)|\Omega\rangle\\
+\prod_{\beta=1}^{M}\frac{g(\Lambda_\beta-u)}{f(\Lambda_\beta-u)}&\left( \prod_{\alpha=1,\alpha\ne \beta}^{M}\frac{1}{f(\Lambda_\beta-\Lambda_\alpha)}-\prod_{\alpha=1,\alpha\ne \beta}^{M}\frac{h(\Lambda_\alpha-\Lambda_\beta)}{f(\Lambda_\alpha-\Lambda_\beta )}\prod_{j=1}^{N}f(k_j-\Lambda_\beta)\right)B(u)\prod_{j=1,j\ne \beta}^{M}B(\Lambda_j)|\Omega\rangle
\end{aligned}
\end{equation}
when
\begin{equation}
\begin{aligned}
t_2&=\prod_{\alpha=1,\alpha\ne \beta}^{M}\frac{1}{f(\Lambda_\beta-\Lambda_\alpha)}-\prod_{\alpha=1,\alpha\ne \beta}^{M}\frac{h(\Lambda_\alpha-\Lambda_\beta)}{f(\Lambda_\alpha-\Lambda_\beta )}\prod_{j=1}^{N}f(k_j-\Lambda_\beta)=0,\\
=&\prod_{\alpha=1,\alpha\ne \beta}^{M}\frac{\Lambda_\beta-\Lambda_\alpha+ic}{\Lambda_\beta-\Lambda_\alpha}-\prod_{\alpha=1,\alpha\ne \beta}^{M}\frac{\Lambda_\beta-\Lambda_\alpha+ic}{\Lambda_\beta-\Lambda_\alpha}\prod_{j=1}^{N}\frac{k_j-\Lambda_\beta}{k_j-\Lambda_\beta+ic}.
\end{aligned}
\end{equation}
The eigenvalue of the transfer matrix is given by 
\begin{equation}
\begin{aligned}
t_1(u)&=\prod_{\alpha=1}^{M}\frac{1}{f(u-\Lambda_\alpha)}-\prod_{\alpha=1}^{M}\frac{h(\Lambda_\alpha-u)}{f(\Lambda_\alpha-u)}\prod_{j=1}^{N}f(k_j-u)\\
=&\prod_{\alpha=1}^{M}\frac{u-\Lambda_\alpha+ic}{u-\Lambda_\alpha}-\prod_{\alpha=1}^{M}\frac{u-\Lambda_\alpha+ic}{u-\Lambda_\alpha}\prod_{j=1}^{N}\frac{u-k_j}{u-k_j-ic}.
\end{aligned}
\end{equation}

So we have the Bethe Ansatz equations 
\begin{equation}
\left\lbrace 
\begin{aligned}
t_1(k_j)&=e^{ik_jL}\\
t_2&=0\\
\end{aligned}
\right.
\end{equation}
which are
\begin{equation}
\begin{aligned}
&\prod_{\alpha=1}^{M}\frac{k_j-\Lambda_\alpha+ic}{k_j-\Lambda_\alpha}=e^{ik_jL},\\
&\prod_{j=1}^{N}\frac{k_j-\Lambda_\beta+ic}{k_j-\Lambda_\beta}=1.\\
\end{aligned}
\end{equation}
Making a shift in $\Lambda$, $\Lambda\to \Lambda+\frac{ic}{2}$, then as same as the results of\cite{imambekov2006exactly,imambekov2006applications}, we obtain the Bethe Ansatz equations
\begin{equation}
\begin{aligned}
&\prod_{\alpha=1}^{M}\frac{k_j-\Lambda_\alpha+\frac{ic}{2}}{k_j-\Lambda_\alpha-\frac{ic}{2}}=e^{ik_jL},\\
&\prod_{j=1}^{N}\frac{k_j-\Lambda_\beta+\frac{ic}{2}}{k_j-\Lambda_\beta-\frac{ic}{2}}=1.\\
\end{aligned}
\end{equation}

\section{II. Generalized hydrodynamics }

The main idea of local equilibrium approximation is that we can separate a system into many fluid  cells, and all these cells are large enough to define the macroscopical properties and small enough to be homogeneous after appropriate relaxation time. Then a state can be characterized by 
\begin{equation}
\hat{\rho}=\hat{\rho}_{x_1}\otimes\hat{\rho}_{x_2}\otimes...\otimes\hat{\rho}_{x_n}\otimes...
\end{equation}
where $\hat{\rho}_{x_i}$ is the density matrix of  equilibrium state and $x_i$ is the position of the $i$th fluid cell. 

For integrable systems, there are many local conserved quantities ${\hat{Q}^1, \hat{Q}^2,...,\hat{Q}^N}$. An equilibrium state is a maximal entropy state constrained by all the conserved quantities. So the density matrix is given by\cite{rigol2007relaxation}
\begin{equation}
\hat{\rho}=\frac{e^{-\sum_{i}\beta^i\hat{Q}^i}}{Tr[e^{-\sum_{i}\beta^i\hat{Q}^i}]}.
\end{equation}
Therefore, an equilibrium state of integrable systems can be identified by the expectation values of all the conserved quantities $Q^1, Q^2,...,Q^N$, where $Q^i=\langle\hat{Q}^i\rangle$.

Analogously, an non-equilibrium state, under local equilibrium approximation, of integrable systems can be identified by the distributions of all the conserved quantities  $q^1(x),q^2(x),...,q^N(x),...$, where $x\in [0,L]$ and $q^i=\langle\hat{q}^i\rangle$,
\begin{equation}
\hat{Q}^i=\int dx \hat{q}^i(x).
\end{equation}

Now let's consider the time  evolution of the system. 
According to the generalized hydrodynamics, we can prove  all the local conserved quantities satisfy the continuity equations. As we have showed in the text, for Bose-Fermi mixture system, there are two independent sets of the conserved quantities, and the continuity equations are given by the followings
\begin{equation}\label{continuityequations}
\begin{aligned}
\partial_tq^i_\rho(x,t)+\partial_xj^i_\rho(x,t)&=0, \\
\partial_tq^i_\sigma(x,t)+\partial_xj^i_\sigma(x,t)&=0, \ i=1,2,...
\end{aligned}
\end{equation}
where 
\begin{equation}
\begin{aligned}
q^i_\rho (x,t)&=\int dk\,\rho (k,x,t)h_\rho^i(k),\\
q^i_\sigma (x,t)&=\int d\Lambda\,\sigma(\Lambda,x,t)h_\sigma^i(\Lambda),\\
j^i_\rho(x,t)&=\int dkv_\rho(k,x,t)\,\rho (k,x,t)h_\rho^i(k),\\
j^i_\sigma(x,t)&=\int d\Lambda v_\sigma (\Lambda,x,t)\,\sigma (\Lambda,x,t)h_\sigma^i(\Lambda).
\end{aligned}
\end{equation}
Then, we have 
\begin{equation}\label{ICE}
\begin{aligned}
\int dk\left( \partial_t \rho(k,x,t)+\partial_x (v_\rho(k,x,t)\rho(k,x,t))\right) h_\rho^i(k)&=0,\\
\int d\Lambda\left( \partial_t \sigma(\Lambda,x,t)+\partial_x (v_\sigma(\Lambda,x,t)\sigma(\Lambda,x,t))\right) h_\sigma^i(\Lambda)&=0.\\
\end{aligned}
\end{equation}
Where $v_\rho$ and $v_\sigma$ are the sound velocities of the excitations.  In the above equations, we denoted 
\begin{equation}\label{sound_velocity}
\begin{aligned}
v_\rho(k)=&\frac{\partial \varepsilon(k)}{\partial p_\rho^{dr}(k)}=\frac{ \varepsilon'(k)}{p_\rho'^{dr}(k)},\\
v_\sigma(\Lambda)=&\frac{\partial \varphi(\Lambda)}{\partial p_\sigma^{dr}(\Lambda)}=\frac{\varphi'(\Lambda)}{p_\sigma'^{dr}(\Lambda)},
\end{aligned}
\end{equation}
where $\varepsilon(k)$ and $\varphi(\Lambda)$ are the energies of the excitations, $p_\rho^{dr}(k)$ and  $p_\sigma^{dr}(\Lambda)$ are the momenta of the excitations.

For quantum integrable systems, when we consider the excited states, the derivatives of momenta and energies have the following forms,
\begin{equation}\label{dre_M}
\begin{aligned}
p_\rho'^{dr}(k)=& 1+\int_{-\infty}^{\infty}d\Lambda a(k,\Lambda)n_\sigma(\Lambda)p_\sigma'^{dr}(\Lambda),\\
p_\sigma'^{dr}(\Lambda)=&\int_{-\infty}^{\infty}dka(\Lambda,k)n_\rho(k)p_\rho'^{dr}(k),\\
\end{aligned}
\end{equation}
\begin{equation}\label{dre_E}
\begin{aligned}
\varepsilon'(k)=& e'(k)+\int_{-\infty}^{\infty}d\Lambda a(k,\Lambda)n_\sigma(\Lambda)\varphi'(\Lambda),\\
\varphi'(\Lambda)=&\int_{-\infty}^{\infty}dka(\Lambda,k)n_\rho(k)\varepsilon'(k).\\
\end{aligned}
\end{equation}
From (\ref{dre_M}), we can easily find that
\begin{equation}\label{dre_M'}
\begin{aligned}
p_\rho'^{dr}(k)=&2\pi[\rho(k)+\rho_h(k)], \\
p_\sigma'^{dr}(\Lambda)=&2\pi[\sigma(\Lambda)+\sigma_h(\Lambda)].
\end{aligned}
\end{equation}

These two equations can give us the numerical results of sound velocities (\ref{sound_velocity}) for given $n_\rho(k)$ and $n_\sigma(\Lambda)$. Here we use them to simplify (\ref{ICE}).
We see that $h_\rho^i(k)$ and $h_\sigma^i(\Lambda)$ are the basis vectors of two function spaces. Therefore the  two differential equations can be obtained from (\ref{ICE})
\begin{equation}
\begin{aligned}
\partial_t \rho(k,x,t)+\partial_x (v_\rho(k,x,t)\rho(k,x,t))&=0,\\
\partial_t \sigma(\Lambda,x,t)+\partial_x (v_\sigma(\Lambda,x,t)\sigma(\Lambda,x,t))&=0.\\
\end{aligned}
\end{equation}
These two equations can be further simplified. The proof  can be found below. 

From (\ref{dre_M'}) and (\ref{sound_velocity}), we have
\begin{equation}
\begin{aligned}
&\partial_t \rho(k,x,t)+\partial_x (v_\rho(k,x,t)\rho(k,x,t))
=\frac{1}{2\pi}\partial_t(n_\rho(k,x,t)p_\rho'^{dr}(k,x,t))+\frac{1}{2\pi}\partial_x(n_\rho(k,x,t)\varepsilon'(k,x,t))\\
=&\frac{1}{2\pi}p_\rho'^{dr}(k,x,t)[\partial_tn_\rho(k,x,t)+v_\rho(k,x,t)\partial_xn_\rho(k,x,t)]+\frac{1}{2\pi}n_\rho(k,x,t)[\partial_tp_\rho'^{dr}(k,x,t)+\partial_x\varepsilon'(k,x,t)]=0,\\
&\partial_t \sigma(\Lambda,x,t)+\partial_x (v_\sigma(\Lambda,x,t)\sigma(\Lambda,x,t))
=\frac{1}{2\pi}\partial_t(n_\sigma(\Lambda,x,t)p_\sigma'^{dr}(\Lambda,x,t))+\frac{1}{2\pi}\partial_x(n_\sigma(\Lambda,x,t)\varphi'(\Lambda,x,t))\\
=&\frac{1}{2\pi}p_\sigma'^{dr}(\Lambda,x,t)[\partial_tn_\sigma(\Lambda,x,t)+v_\sigma(\Lambda,x,t)\partial_xn_\sigma(\Lambda,x,t)]+\frac{1}{2\pi}n_\sigma(\Lambda,x,t)[\partial_tp_\sigma'^{dr}(\Lambda,x,t)+\partial_x\varphi'(\Lambda,x,t)]=0.\\
\end{aligned}
\end{equation}
From (\ref{dre_M}) and (\ref{dre_E}), we have the relations 
\begin{equation}
\begin{aligned}
&\partial_tp_\rho'^{dr}(k,x,t)+\partial_x\varepsilon'(k,x,t)=\int d\Lambda a(k,\Lambda)[\partial_t(n_\sigma(\Lambda,x,t)p_\sigma'^{dr}(\Lambda,x,t))+\partial_x(n_\sigma(\Lambda,x,t)\varphi'(\Lambda,x,t))]=0, \\
&\partial_tp_\sigma'^{dr}(\Lambda,x,t)+\partial_x\varphi'(\Lambda,x,t)=\int dka(\Lambda,k)[\partial_t(n_\rho(k,x,t)p_\rho'^{dr}(k,x,t))+\partial_x(n_\rho(k,x,t)\varepsilon'(k,x,t))]=0. 
\end{aligned}
\end{equation}
Finally, as same as the results of \cite{castro2016emergent,bertini2016transport,bertini2018low,doyon2017note,mestyan2019spin,bulchandani2018bethe,bulchandani2017solvable}
\begin{equation}\label{SICE}
\begin{aligned}
&\partial_tn_\rho(k,x,t)+v_\rho(k,x,t)\partial_xn_\rho(k,x,t)=0, \\
&\partial_tn_\sigma(\Lambda,x,t)+v_\sigma(\Lambda,x,t)\partial_xn_\sigma(\Lambda,x,t)=0.
\end{aligned}
\end{equation}

\section{III. Analytical calculation of the quantum transport properties}
Now, let's work on the low-temperature transport. For the transport propertities, we will show that the description of generalized hydrodynamics is truly equivalent to two decoupled Luttinger liquids\cite{guan2013fermi}. The foundation of low-temperature expansion is  the Sommerfeld expansion. After the expansion, all the results can be expressed by the polynomials in terms of the temperatures, and all the coefficients are related to the ground state \cite{bertini2018low,bertini2018universal,mestyan2019spin}.

Let's introduce the dressed charges $f_q(k)$ and $g_q(\Lambda)$  which are the charges carried by the excitations
\begin{equation}\label{dressedq}
\begin{aligned}
f_q(k)=&f(k)+\int_{-\Lambda_0}^{\Lambda_0}a(k-\Lambda)g_q(\Lambda)d\Lambda,\\
g_q(\Lambda)=&g(\Lambda)+\int_{-k_0}^{k_0}a(\Lambda-k)f_q(k)dk,
\end{aligned}
\end{equation}
where $f(k)$ and $g(\Lambda)$ are the bare energies of  single particles. For example,  we consider the energies of the excitations, $f(k)=k^2-\mu_f$ and $g(\Lambda)=\mu_f-\mu_b$ in charge and bosonic degrees of freedom.

From the expressions of the densities and currents,
\begin{equation}
\begin{aligned}
q(\xi)=&\int_{-\infty}^{\infty}f(k)n_\rho(k,\xi)\rho^t(k,\xi)dk+\int_{-\infty}^{\infty}g(\Lambda)n_\sigma(\Lambda,\xi)\sigma^t(\Lambda,\xi)d\Lambda\\
j(\xi)=&\int_{-\infty}^{\infty}f(k)n_\rho(k,\xi)v_\rho(k,\xi)\rho^t(k,\xi)dk+\int_{-\infty}^{\infty}g(\Lambda)n_\sigma(\Lambda,\xi)v_\sigma(\Lambda,\xi)\sigma^t(\Lambda,\xi)d\Lambda\\
\end{aligned}
\end{equation} 

For equilibrium state,
\begin{equation}
\begin{aligned}
&n_\rho(k)=\frac{\rho(k)}{\rho^t(k)}=\frac{1}{e^{\beta\varepsilon(k)}+1},\\
&n_\sigma(\Lambda)=\frac{\sigma(\Lambda)}{\sigma^t(\Lambda)}=\frac{1}{e^{\beta\varphi(\Lambda)}+1}.
\end{aligned}
\end{equation}
We find that  the main calculation of the hydrodynamic will involve the expressions of the densities and currents like 
\begin{equation}
\int_{-\infty}^{\infty}dkn_\rho(k)p(k),\ \int_{-\infty}^{\infty}d\Lambda n_\sigma(\Lambda)q(\Lambda). 
\end{equation}
Then  we can expand these by the  Sommerfeld expansion
\begin{equation}\label{SE}
\int_{0}^{\infty}\frac{p(k)}{e^{\beta\varepsilon(k)}+1}dk=\int_{0}^{k'_0}dk p(k)+\frac{\pi^2T^2}{6}\frac{p'(k'_0)}{(\varepsilon'(k'_0))^2}-\frac{\pi^2T^2}{6}\frac{\varepsilon''(k'_0)p(k_0)}{(\varepsilon'(k'_0))^3}+O(T^4)\quad(\varepsilon(k'_0)=0;T\to 0).
\end{equation}
We have the same equation as (\ref{SE}) for $\int_{-\infty}^{\infty}d\Lambda n_\sigma(\Lambda)q(\Lambda)$. Here, $k'_0$ is related  to the temperature, but what we want to achieve is that all the  coefficients of the polynomials are given only by the ground state properties. So we need to find the relations between $k'_0$ and $k_0$.

From the thermodynamic Bethe Ansatz equations, we have
\begin{equation}\label{Tdrek}
\begin{aligned}
&\varepsilon(k)-\varepsilon_0(k)=-T\int_{-\infty}^{\infty}a(k-\Lambda)\ln(1+e^{-\varphi(\Lambda)/T})d\Lambda-\int_{-\Lambda_0}^{\Lambda_0}a(k-\Lambda)\varphi_0(\Lambda)d\Lambda\\
=&\int_{-\Lambda_0}^{\Lambda_0}a(k-\Lambda)(\varphi(\Lambda)-\varphi_0(\Lambda))d\Lambda+\int_{\Lambda_0}^{\Lambda_0'}a(k-\Lambda)\varphi(\Lambda)d\Lambda+\int_{-\Lambda_0'}^{-\Lambda_0}a(k-\Lambda)\varphi(\Lambda)d\Lambda\\
&-T\int_{-\infty}^{\infty}a(k-\Lambda)\ln(1+e^{-|\varphi(\Lambda)|/T})d\Lambda\quad(\varphi(\pm\Lambda_0')=0)\\
\approx&\int_{-\Lambda_0}^{\Lambda_0}a(k-\Lambda)(\varphi(\Lambda)-\varphi_0(\Lambda))d\Lambda-\frac{\pi^2T^2}{6\varphi_0'(\Lambda_0)}\left( a(k-\Lambda_0)+a(k+\Lambda_0)\right). 
\end{aligned}
\end{equation}
Samely, we have
\begin{equation}\label{TdreL}
\varphi(\Lambda)-\varphi_0(\Lambda)=\int_{-k_0}^{k_0}a(\Lambda-k)(\varepsilon(k)-\varepsilon_0(k))dk-\frac{\pi^2T^2}{6\varepsilon_0'(k_0)}\left( a(\Lambda-k_0)+a(\Lambda+k_0)\right).
\end{equation}
We then  define the following relations  to simplify the above expressions
\begin{equation}\label{AB}
\begin{aligned}
\varepsilon(k)-\varepsilon_0(k)=&\frac{\pi^2T^2}{6\varepsilon_0'(k_0)}A(k),\\
\varphi(\Lambda)-\varphi_0(\Lambda)=&\frac{\pi^2T^2}{6\varphi_0'(\Lambda_0)}B(\Lambda).\\
\end{aligned}
\end{equation}
Then, (\ref{Tdrek}) and (\ref{TdreL}) become
\begin{equation}
\begin{aligned}
\frac{\varphi_0'(\Lambda_0)}{\varepsilon_0'(k_0)}A(k)=&\int_{-\Lambda_0}^{\Lambda_0}a(k-\Lambda)B(\Lambda)d\Lambda-\left( a(k-\Lambda_0)+a(k+\Lambda_0)\right),\\
\frac{\varepsilon_0'(k_0)}{\varphi_0'(\Lambda_0)}B(\Lambda)=&\int_{-k_0}^{k_0}a(\Lambda-k)A(k)dk-\left( a(\Lambda-k_0)+a(\Lambda+k_0)\right).\\
\end{aligned}
\end{equation}
From (\ref{AB}), we obtain what we need 
\begin{equation}
\begin{aligned}
k'_0-k_0=&-\frac{\pi^2T^2}{6(\varepsilon_0'(k_0))^2}A(k_0),\\
\Lambda'_0-\Lambda_0=&-\frac{\pi^2T^2}{6(\varphi_0'(\Lambda_0))^2}B(\Lambda_0).
\end{aligned}
\end{equation}
Finally, we obtain the proper form of (\ref{SE}),
\begin{equation}
\int_{-\infty}^{\infty}n_\rho (k)p(k)dk=\int_{-k_0}^{k_0}p(k)dk+\frac{\pi^2T^2}{6(\varepsilon_0'(k_0))^2}\left[ p'(k_0)-p'(-k_0)-\left( \frac{\varepsilon''_0(k_0)}{\varepsilon'_0(k_0)}+A(k_0)\right) \left( p(k_0)+p(-k_0)\right) \right]. 
\end{equation}
Samely for $\int_{-\infty}^{\infty}d\Lambda n_\sigma(\Lambda)q(\Lambda)$,
\begin{equation}
\int_{-\infty}^{\infty}n_\sigma (\Lambda)q(\Lambda)d\Lambda=\int_{-\Lambda_0}^{\Lambda_0}q(\Lambda)d\Lambda+\frac{\pi^2T^2}{6(\varphi_0'(\Lambda_0))^2}\left[ q'(\Lambda_0)-q'(-\Lambda_0)-\left( \frac{\varphi''_0(\Lambda_0)}{\varphi'_0(\Lambda_0)}+B(\Lambda_0)\right) \left( q(\Lambda_0)+q(-\Lambda_0)\right) \right].
\end{equation}

Now, let's consider the low-temperature expansion for non-equilibrium state with the given intial state which is  set up as two semi-infinite halves with different temperatures joined together at $t=0$, then the solutions of (\ref{SICE}) are given by 
\begin{eqnarray}
	n_\rho(k,\xi)=n^L_\rho(k)H(v_\rho(k,\xi)-\xi)+n^R_\rho(k)H(\xi-v_\rho(k,\xi)),\nonumber \\
	n_\sigma(\Lambda,\xi)=n^L_\sigma(\Lambda)H(v_\sigma(\Lambda,\xi)-\xi)+n^R_\sigma(\Lambda)H(\xi-v_\sigma(\Lambda,\xi)).  \label{final_solutions}
\end{eqnarray}

First, we consider the region $\lim\limits_{T\to0}|\xi\pm v^0_\rho(k_0)|\ne0$ and $\lim\limits_{T\to0}|\xi\pm v^0_\sigma(\Lambda_0)|\ne0$, where there is no singularity for the Heaviside step functions in (\ref{final_solutions}) at zero temperature, so, the expansions will be  performed 
\begin{equation}\label{firstexpansion}
\begin{aligned}
&q(\xi)=\int_{-\infty}^{\infty}f(k)n_\rho(k,\xi)\rho^t(k,\xi)dk+\int_{-\infty}^{\infty}g(\Lambda)n_\sigma(\Lambda,\xi)\sigma^t(\Lambda,\xi)d\Lambda\\
=&q_0+\int_{-k_0}^{k_0}f(k)(\rho^t(k,\xi)-\rho_0^t(k))dk+\int_{-\Lambda_0}^{\Lambda_0}g(\Lambda)(\sigma^t(\Lambda,\xi)-\sigma_0^t(\Lambda))d\Lambda\\
&+\frac{\pi T_L^2}{12\varepsilon_0'(k_0)v^0_\rho(k_0)}\left( f'(k_0)-\left( \frac{v^0_\rho{'}(k_0)}{v^0_\rho(k_0)}+A(k_0)\right) f(k_0)\right) H(v^0_\rho(k_0)-\xi) \\
&+\frac{\pi T_L^2}{12\varepsilon_0'(k_0)v^0_\rho(k_0)}\left( -f'(-k_0)-\left( \frac{v^0_\rho{'}(k_0)}{v^0_\rho(k_0)}+A(k_0)\right) f(-k_0)\right)H(-v^0_\rho(k_0)-\xi) \\
&+\frac{\pi T_R^2}{12\varepsilon_0'(k_0)v^0_\rho(k_0)}\left( f'(k_0)-\left( \frac{v^0_\rho{'}(k_0)}{v^0_\rho(k_0)}+A(k_0)\right) f(k_0)\right) H(\xi-v^0_\rho(k_0)) \\
&+\frac{\pi T_R^2}{12\varepsilon_0'(k_0)v^0_\rho(k_0)}\left( -f'(-k_0)-\left( \frac{v^0_\rho{'}(k_0)}{v^0_\rho(k_0)}+A(k_0)\right) f(-k_0)\right)H(\xi+v^0_\rho(k_0))\\
&+\frac{\pi T_L^2}{12\varphi_0'(\Lambda_0)v^0_\sigma(\Lambda_0)}\left( g'(\Lambda_0)-\left(\frac{v^0_\sigma{'}(\Lambda_0)}{v^0_\sigma(\Lambda_0)}+B(\Lambda_0)\right) g(\Lambda_0)\right)H(v^0_\sigma (\Lambda_0)-\xi)\\
&+\frac{\pi T_L^2}{12\varphi_0'(\Lambda_0)v^0_\sigma(\Lambda_0)}\left( -g'(-\Lambda_0)-\left(\frac{v^0_\sigma{'}(\Lambda_0)}{v^0_\sigma(\Lambda_0)}+B(\Lambda_0)\right) g(-\Lambda_0)\right)H(-v^0_\sigma(\Lambda_0)-\xi)\\
&+\frac{\pi T_R^2}{12\varphi_0'(\Lambda_0)v^0_\sigma(\Lambda_0)}\left( g'(\Lambda_0)-\left(\frac{v^0_\sigma{'}(\Lambda_0)}{v^0_\sigma(\Lambda_0)}+B(\Lambda_0)\right) g(\Lambda_0)\right)H(\xi-v^0_\sigma (\Lambda_0))\\
&+\frac{\pi T_R^2}{12\varphi_0'(\Lambda_0)v^0_\sigma(\Lambda_0)}\left( -g'(-\Lambda_0)-\left(\frac{v^0_\sigma{'}(\Lambda_0)}{v^0_\sigma(\Lambda_0)}+B(\Lambda_0)\right) g(-\Lambda_0)\right)H(\xi+v^0_\sigma(\Lambda_0)). \\
\end{aligned}
\end{equation}
 The terms $\rho^t(k,\xi)-\rho_0^t(k)$ and $\sigma^t(\Lambda,\xi)-\sigma_0^t(\Lambda)$ in (\ref{firstexpansion}) are needed to calculated from the Bethe Ansatz equations at  finite and zero temperatures, namely 
\begin{equation}\label{tBAEk}
\begin{aligned}
\rho^t(k,\xi)-\rho_0^t(k)=&\int_{-\infty}^{\infty}a(k-\Lambda)n_\sigma(\Lambda,\xi)\sigma^t(\Lambda,\xi)d\Lambda-\int_{-\Lambda_0}^{\Lambda_0}a(k-\Lambda)\sigma_0^t(\Lambda)d\Lambda
=\int_{-\Lambda_0}^{\Lambda_0}a(k-\Lambda)(\sigma^t(\Lambda,\xi)-\sigma_0^t(\Lambda))d\Lambda\\
+&\frac{\pi T_L^2}{12\varphi_0'(\Lambda_0)v^0_\sigma(\Lambda_0)}\left(a'(k-\Lambda_0)-\left(\frac{v^0_\sigma{'}(\Lambda_0)}{v^0_\sigma(\Lambda_0)}+B(\Lambda_0) \right)a(k-\Lambda_0) \right) H(v^0_\sigma (\Lambda_0)-\xi)\\
+&\frac{\pi T_L^2}{12\varphi_0'(\Lambda_0)v^0_\sigma(\Lambda_0)}\left(-a'(k+\Lambda_0)-\left(\frac{v^0_\sigma{'}(\Lambda_0)}{v^0_\sigma(\Lambda_0)}+B(\Lambda_0) \right)a(k+\Lambda_0) \right) H(-v^0_\sigma (\Lambda_0)-\xi)\\
+&\frac{\pi T_R^2}{12\varphi_0'(\Lambda_0)v^0_\sigma(\Lambda_0)}\left(a'(k-\Lambda_0)-\left(\frac{v^0_\sigma{'}(\Lambda_0)}{v^0_\sigma(\Lambda_0)}+B(\Lambda_0) \right)a(k-\Lambda_0) \right) H(\xi-v^0_\sigma (\Lambda_0))\\
+&\frac{\pi T_R^2}{12\varphi_0'(\Lambda_0)v^0_\sigma(\Lambda_0)}\left(-a'(k+\Lambda_0)-\left(\frac{v^0_\sigma{'}(\Lambda_0)}{v^0_\sigma(\Lambda_0)}+B(\Lambda_0) \right)a(k+\Lambda_0) \right) H(\xi+v^0_\sigma (\Lambda_0)),\\
\end{aligned}
\end{equation}
\begin{equation}\label{tBAEL}
\begin{aligned}
\sigma^t(\Lambda,\xi)-\sigma_0^t(\Lambda)=&\int_{-\infty}^{\infty}a(\Lambda-k)n_\rho(k,\xi)\rho^t(k,\xi)dk-\int_{-k_0}^{k_0}a(\Lambda-k)\rho_0^t(k)dk
=\int_{-k_0}^{k_0}a(\Lambda-k)(\rho^t(k,\xi)-\rho_0^t(k))dk\\
+&\frac{\pi T_L^2}{12\varepsilon_0'(k_0)v^0_\rho(k_0)}\left(a'(\Lambda-k_0)-\left(\frac{v^0_\rho{'}(k_0)}{v^0_\rho(k_0)}+A(k_0) \right)a(\Lambda-k_0) \right) H(v^0_\rho (k_0)-\xi)\\
+&\frac{\pi T_L^2}{12\varepsilon_0'(k_0)v^0_\rho(k_0)}\left(-a'(\Lambda+k_0)-\left(\frac{v^0_\rho{'}(k_0)}{v^0_\rho(k_0)}+A(k_0) \right)a(\Lambda+k_0) \right) H(-v^0_\rho(k_0)-\xi)\\
+&\frac{\pi T_R^2}{12\varepsilon_0'(k_0)v^0_\rho(k_0)}\left(a'(\Lambda-k_0)-\left(\frac{v^0_\rho{'}(k_0)}{v^0_\rho(k_0)}+A(k_0) \right)a(\Lambda-k_0) \right) H(\xi-v^0_\rho (k_0))\\
+&\frac{\pi T_R^2}{12\varepsilon_0'(k_0)v^0_\rho(k_0)}\left(-a'(\Lambda+k_0)-\left(\frac{v^0_\rho{'}(k_0)}{v^0_\rho(k_0)}+A(k_0) \right)a(\Lambda+k_0) \right) H(\xi+v^0_\rho (k_0)).\\
\end{aligned}
\end{equation}
After multiplication and integration of (\ref{tBAEk})(\ref{tBAEL})(\ref{dressedq}), we have 
\begin{equation}
\begin{aligned}
&\int_{-k_0}^{k_0}f(k)(\rho^t(k,\xi)-\rho_0^t(k))dk+\int_{-\Lambda_0}^{\Lambda_0}g(\Lambda)(\sigma^t(\Lambda,\xi)-\sigma_0^t(\Lambda))d\Lambda\\
=&\frac{\pi T_L^2}{12\varphi_0'(\Lambda_0)v^0_\sigma(\Lambda_0)}\left((g'_q(\Lambda_0)-g'(\Lambda_0))-\left(\frac{v^0_\sigma{'}(\Lambda_0)}{v^0_\sigma(\Lambda_0)}+B(\Lambda_0) \right)(g_q(\Lambda_0)-g(\Lambda_0)) \right) H(v^0_\sigma (\Lambda_0)-\xi)\\
&+\frac{\pi T_L^2}{12\varphi_0'(\Lambda_0)v^0_\sigma(\Lambda_0)}\left(-(g'_q(-\Lambda_0)-g'(-\Lambda_0))-\left(\frac{v^0_\sigma{'}(\Lambda_0)}{v^0_\sigma(\Lambda_0)}+B(\Lambda_0) \right)(g_q(-\Lambda_0)-g(-\Lambda_0)) \right) H(-v^0_\sigma (\Lambda_0)-\xi)\\
&+\frac{\pi T_R^2}{12\varphi_0'(\Lambda_0)v^0_\sigma(\Lambda_0)}\left((g'_q(\Lambda_0)-g'(\Lambda_0))-\left(\frac{v^0_\sigma{'}(\Lambda_0)}{v^0_\sigma(\Lambda_0)}+B(\Lambda_0) \right)(g_q(\Lambda_0)-g(\Lambda_0)) \right) H(\xi-v^0_\sigma (\Lambda_0))\\
&+\frac{\pi T_R^2}{12\varphi_0'(\Lambda_0)v^0_\sigma(\Lambda_0)}\left(-(g'_q(-\Lambda_0)-g'(-\Lambda_0))-\left(\frac{v^0_\sigma{'}(\Lambda_0)}{v^0_\sigma(\Lambda_0)}+B(\Lambda_0) \right)(g_q(-\Lambda_0)-g(-\Lambda_0)) \right) H(\xi+v^0_\sigma (\Lambda_0))\\
&+\frac{\pi T_L^2}{12\varepsilon_0'(k_0)v^0_\rho(k_0)}\left((f'_q(k_0)-f'(k_0))-\left(\frac{v^0_\rho{'}(k_0)}{v^0_\rho(k_0)}+A(k_0) \right)(f_q(k_0)-f(k_0)) \right) H(v^0_\rho (k_0)-\xi)\\
&+\frac{\pi T_L^2}{12\varepsilon_0'(k_0)v^0_\rho(k_0)}\left(-(f'_q(-k_0)-f'(-k_0))-\left(\frac{v^0_\rho{'}(k_0)}{v^0_\rho(k_0)}+A(k_0) \right)(f_q(-k_0)-f(-k_0)) \right) H(-v^0_\rho(k_0)-\xi)\\
&+\frac{\pi T_R^2}{12\varepsilon_0'(k_0)v^0_\rho(k_0)}\left((f'_q(k_0)-f'(k_0))-\left(\frac{v^0_\rho{'}(k_0)}{v^0_\rho(k_0)}+A(k_0) \right)(f_q(k_0)-f(k_0)) \right) H(\xi-v^0_\rho (k_0))\\
&+\frac{\pi T_R^2}{12\varepsilon_0'(k_0)v^0_\rho(k_0)}\left(-(f'_q(-k_0)-f'(-k_0))-\left(\frac{v^0_\rho{'}(k_0)}{v^0_\rho(k_0)}+A(k_0) \right)(f_q(-k_0)-f(-k_0)) \right) H(\xi+v^0_\rho (k_0)).\\
\end{aligned}
\end{equation}

Finally, we have 
\begin{equation}
\begin{aligned}
q(\xi)=&q_0+\frac{\pi }{12\varphi_0'(\Lambda_0)v^0_\sigma(\Lambda_0)}\left(T_L^2g'_q(\Lambda_0)-T_R^2g'_q(-\Lambda_0)-(T_L^2g_q(\Lambda_0)+T_R^2g_q(-\Lambda_0))\left(\frac{v^0_\sigma{'}(\Lambda_0)}{v^0_\sigma(\Lambda_0)}+B(\Lambda_0) \right) \right) H(v^0_\sigma (\Lambda_0)-|\xi|)\\
&+\frac{\pi }{12\varphi_0'(\Lambda_0)v^0_\sigma(\Lambda_0)}\left(T_L^2g'_q(\Lambda_0)-T_L^2g'_q(-\Lambda_0)-(T_L^2g_q(\Lambda_0)+T_L^2g_q(-\Lambda_0))\left(\frac{v^0_\sigma{'}(\Lambda_0)}{v^0_\sigma(\Lambda_0)}+B(\Lambda_0) \right) \right) H(-v^0_\sigma (\Lambda_0)-\xi)\\
&+\frac{\pi }{12\varphi_0'(\Lambda_0)v^0_\sigma(\Lambda_0)}\left(T_R^2g'_q(\Lambda_0)-T_R^2g'_q(-\Lambda_0)-(T_R^2g_q(\Lambda_0)+T_R^2g_q(-\Lambda_0))\left(\frac{v^0_\sigma{'}(\Lambda_0)}{v^0_\sigma(\Lambda_0)}+B(\Lambda_0) \right) \right) H(\xi-v^0_\sigma (\Lambda_0))\\
&+\frac{\pi }{12\varepsilon_0'(k_0)v^0_\rho(k_0)}\left(T_L^2f'_q(k_0)-T_R^2f'_q(-k_0)-(T_L^2f_q(k_0)+T_R^2f_q(-k_0))\left(\frac{v^0_\rho{'}(k_0)}{v^0_\rho(k_0)}+A(k_0) \right) \right) H(v^0_\rho (k_0)-|\xi|)\\
&+\frac{\pi }{12\varepsilon_0'(k_0)v^0_\rho(k_0)}\left(T_L^2f'_q(k_0)-T_L^2f'_q(-k_0)-(T_L^2f_q(k_0)+T_L^2f_q(-k_0))\left(\frac{v^0_\rho{'}(k_0)}{v^0_\rho(k_0)}+A(k_0) \right) \right) H(-v^0_\rho(k_0)-\xi)\\
&+\frac{\pi }{12\varepsilon_0'(k_0)v^0_\rho(k_0)}\left(T_R^2f'_q(k_0)-T_R^2f'_q(-k_0)-(T_R^2f_q(k_0)+T_R^2f_q(-k_0))\left(\frac{v^0_\rho{'}(k_0)}{v^0_\rho(k_0)}+A(k_0) \right) \right) H(\xi-v^0_\rho (k_0)). 
\end{aligned}
\end{equation}

Now, we have the low-temperature expansion for the densities of the conserved quantities. For the dressed  energies 
\begin{equation}\label{zeroTBAE}
\begin{aligned}
f_e(k_0)=\varepsilon_0(k)=0, &\\
g_e(\Lambda_0)=\varphi_0(\Lambda)=0.&\\
\end{aligned}
\end{equation}
Therefore, we have 
\begin{equation}
\begin{aligned}
q_e(\xi)=e_0&+\frac{\pi }{12v^0_\sigma(\Lambda_0)}\left(T_L^2+T_R^2\right) H(v^0_\sigma (\Lambda_0)-|\xi|)+\frac{\pi T_L^2 }{6v^0_\sigma(\Lambda_0)} H(-v^0_\sigma (\Lambda_0)-\xi)+\frac{\pi T_R^2 }{6v^0_\sigma(\Lambda_0)} H(\xi-v^0_\sigma (\Lambda_0))\\
&+\frac{\pi }{12v^0_\rho(k_0)}\left(T_L^2+T_R^2\right) H(v^0_\rho (k_0)-|\xi|)+\frac{\pi T_L^2 }{6v^0_\rho(k_0)} H(-v^0_\rho(k_0)-\xi)+\frac{\pi T_R^2 }{6v^0_\rho(k_0)} H(\xi-v^0_\rho (k_0)).\\
\end{aligned}
\end{equation}

Let's move on to the currents
\begin{equation}
\begin{aligned}
&j(\xi)=\int_{-\infty}^{\infty}f(k)n_\rho(k,\xi)v_\rho(k,\xi)\rho^t(k,\xi)dk+\int_{-\infty}^{\infty}g(\Lambda)n_\sigma(\Lambda,\xi)v_\sigma(\Lambda,\xi)\sigma^t(\Lambda,\xi)d\Lambda\\
=&\int_{-k_0}^{k_0}f(k)(v_\rho(k,\xi)\rho^t(k,\xi)-v^0_\rho(k)\rho_0^t(k))dk+\int_{-\Lambda_0}^{\Lambda_0}g(\Lambda)(v_\sigma(\Lambda,\xi)\sigma^t(\Lambda,\xi)-v^0_\sigma(\Lambda)\sigma_0^t(\Lambda))d\Lambda\\
&+\frac{\pi T_L^2}{12\varepsilon_0'(k_0)}\left( f'(k_0)-A(k_0) f(k_0)\right) H(v^0_\rho(k_0)-\xi)+\frac{\pi T_L^2}{12\varepsilon_0'(k_0)}\left( f'(-k_0)+A(k_0) f(-k_0)\right)H(-v^0_\rho(k_0)-\xi) \\
&+\frac{\pi T_R^2}{12\varepsilon_0'(k_0)}\left(f'(k_0)-A(k_0) f(k_0)\right)H(\xi-v^0_\rho(k_0))+\frac{\pi T_R^2}{12\varepsilon_0'(k_0)}\left(f'(-k_0)+A(k_0) f(-k_0)\right)H(\xi+v^0_\rho(k_0)) \\
&+\frac{\pi T_L^2}{12\varphi_0'(\Lambda_0)}\left( g'(\Lambda_0)-B(\Lambda_0)g(\Lambda_0)\right)H(v^0_\sigma (\Lambda_0)-\xi)+\frac{\pi T_L^2}{12\varphi_0'(\Lambda_0)}\left( g'(-\Lambda_0)+B(\Lambda_0)g(-\Lambda_0)\right)H(-v^0_\sigma(\Lambda_0)-\xi)\\
&+\frac{\pi T_R^2}{12\varphi_0'(\Lambda_0)}\left( g'(\Lambda_0)-B(\Lambda_0) g(\Lambda_0)\right)H( \xi-v^0_\sigma(\Lambda_0))+\frac{\pi T_R^2}{12\varphi_0'(\Lambda_0)}\left( g'(-\Lambda_0)+B(\Lambda_0) g(-\Lambda_0)\right)H(\xi+v^0_\sigma(\Lambda_0)).\\
\end{aligned}
\end{equation}

Just like the expansion of densities, now we need to calculate $v_\rho(k,\xi)\rho^t(k,\xi)-v^0_\rho(k)\rho_0^t(k)$ and $v_\sigma(\Lambda,\xi)\sigma^t(\Lambda,\xi)-v^0_\sigma(\Lambda)\sigma_0^t(\Lambda)$. From (\ref{sound_velocity}) (\ref{dre_E}) and (\ref{dre_M'}), we have
\begin{equation}\label{vdre}
\begin{aligned}
v_\rho(k,\xi)\rho^t(k,\xi)=\frac{\varepsilon'(k,\xi)}{2\pi}=&\frac{e'(k)}{2\pi}+\int_{-\infty}^{\infty}a(k-\Lambda)n_\sigma(\Lambda,
\xi)v_\sigma(\Lambda,\xi)\sigma^t(\Lambda,\xi)d\Lambda,\\
v_\sigma(\Lambda,\xi)\sigma^t(\Lambda,\xi)=\frac{\varphi'(\Lambda,\xi)}{2\pi}=&\frac{g'(\Lambda)}{2\pi}+\int_{-\infty}^{\infty}a(\Lambda-k)n_\rho(k,\xi)v_\rho(k,\xi)\rho^t(k,\xi)dk.\\
\end{aligned}
\end{equation}
For ground state, we have (\ref{vdre}) like the following form
\begin{equation}
\begin{aligned}
v^0_\rho(k)\rho_0^t(k)=\frac{\varepsilon'_0(k)}{2\pi}=&\frac{e'(k)}{2\pi}+\int_{-\Lambda_0}^{\Lambda_0}a(k-\Lambda)v^0_\sigma(\Lambda)\sigma_0^t(\Lambda)d\Lambda,\\
v^0_\sigma(\Lambda)\sigma_0^t(\Lambda)=\frac{\varphi'_0(\Lambda)}{2\pi}=&\frac{g'(\Lambda)}{2\pi}+\int_{-k_0}^{k_0}a(\Lambda-k)v^0_\rho(k)\rho_0^t(k)dk.
\end{aligned}
\end{equation}
Then 
\begin{equation}\label{deriTBA}
\begin{aligned}
&v_\rho(k,\xi)\rho^t(k,\xi)-v^0_\rho(k)\rho_0^t(k)=\int_{-\infty}^{\infty}a(k-\Lambda)n_\sigma(\Lambda,
\xi)v_\sigma(\Lambda,\xi)\sigma^t(\Lambda,\xi)d\Lambda-\int_{-\Lambda_0}^{\Lambda_0}a(k-\Lambda)v^0_\sigma(\Lambda)\sigma_0^t(\Lambda)d\Lambda\\
=&\int_{-\Lambda_0}^{\Lambda_0}a(k-\Lambda)(v_\sigma(\Lambda,\xi)\sigma^t(\Lambda,\xi)-v^0_\sigma(\Lambda)\sigma_0^t(\Lambda))d\Lambda+\frac{\pi T_L^2}{12\varphi_0'(\Lambda_0)}\left(a'(k-\Lambda_0)-B(\Lambda_0)a(k-\Lambda_0) \right) H(v^0_\sigma (\Lambda_0)-\xi),\\
+&\frac{\pi T_L^2}{12\varphi_0'(\Lambda_0)}\left(a'(k+\Lambda_0)+B(\Lambda_0) a(k+\Lambda_0)\right) H(-v^0_\sigma (\Lambda_0)-\xi)+\frac{\pi T_R^2}{12\varphi_0'(\Lambda_0)}\left(a'(k-\Lambda_0)-B(\Lambda_0) a(k-\Lambda_0) \right) H(\xi-v^0_\sigma (\Lambda_0))\\
+&\frac{\pi T_R^2}{12\varphi_0'(\Lambda_0)}\left(a'(k+\Lambda_0)+B(\Lambda_0) a(k+\Lambda_0) \right) H(\xi+v^0_\sigma (\Lambda_0)).\\
\end{aligned}
\end{equation}
\begin{equation}\label{deriTBAz}
\begin{aligned}
&v_\sigma(\Lambda,\xi)\sigma^t(\Lambda,\xi)-v^0_\sigma(\Lambda)\sigma_0^t(\Lambda)=\int_{-\infty}^{\infty}a(\Lambda-k)n_\rho(k,
\xi)v_\rho(k,\xi)\rho^t(k,\xi)dk-\int_{-k_0}^{k_0}a(\Lambda-k)v^0_\rho(k)\rho_0^t(k)dk\\
=&\int_{-k_0}^{k_0}a(\Lambda-k)(v_\rho(k,\xi)\rho^t(k,\xi)-v^0_\rho(k)\rho_0^t(k))dk+\frac{\pi T_L^2}{12\varepsilon_0'(k_0)}\left(a'(\Lambda-k_0)-A(k_0) a(\Lambda-k_0) \right) H(v^0_\rho (k_0)-\xi)\\
+&\frac{\pi T_L^2}{12\varepsilon_0'(k_0)}\left(a'(\Lambda+k_0)+A(k_0) a(\Lambda+k_0) \right) H(-v^0_\rho(k_0)-\xi)+\frac{\pi T_R^2}{12\varepsilon_0'(k_0)}\left(a'(\Lambda-k_0)-A(k_0) a(\Lambda-k_0) \right) H(\xi-v^0_\rho (k_0))\\
+&\frac{\pi T_R^2}{12\varepsilon_0'(k_0)}\left(a'(\Lambda+k_0)+A(k_0) a(\Lambda+k_0) \right) H(\xi+v^0_\rho (k_0)).
\end{aligned}
\end{equation}
After multiplication and integration of (\ref{deriTBA})(\ref{deriTBAz})(\ref{dressedq}), we have 
\begin{equation}
\begin{aligned}
&\int_{-k_0}^{k_0}f(k)(v_\rho(k,\xi)\rho^t(k,\xi)-v^0_\rho(k)\rho_0^t(k))dk+\int_{-\Lambda_0}^{\Lambda_0}g(\Lambda)(v_\sigma(\Lambda,\xi)\sigma^t(\Lambda,\xi)-v^0_\sigma(\Lambda)\sigma_0^t(\Lambda))d\Lambda\\
&=\frac{\pi T_L^2}{12\varphi_0'(\Lambda_0)}\left((g'_q(\Lambda_0)-g'(\Lambda_0))-B(\Lambda_0) (g_q(\Lambda_0)-g(\Lambda_0)) \right) H(v^0_\sigma (\Lambda_0)-\xi)\\
&+\frac{\pi T_L^2}{12\varphi_0'(\Lambda_0)}\left((g'_q(-\Lambda_0)-g'(-\Lambda_0))+B(\Lambda_0)(g_q(-\Lambda_0)-g(-\Lambda_0)) \right) H(-v^0_\sigma (\Lambda_0)-\xi)\\
&+\frac{\pi T_R^2}{12\varphi_0'(\Lambda_0)}\left((g'_q(\Lambda_0)-g'(\Lambda_0))-B(\Lambda_0) (g_q(\Lambda_0)-g(\Lambda_0)) \right) H(\xi-v^0_\sigma (\Lambda_0))\\
&+\frac{\pi T_R^2}{12\varphi_0'(\Lambda_0)}\left((g'_q(-\Lambda_0)-g'(-\Lambda_0))+B(\Lambda_0)(g_q(-\Lambda_0)-g(-\Lambda_0)) \right) H(\xi+v^0_\sigma (\Lambda_0))\\
&+\frac{\pi T_L^2}{12\varepsilon_0'(k_0)}\left((f'_q(k_0)-f'(k_0))-A(k_0) (f_q(k_0)-f(k_0)) \right) H(v^0_\rho (k_0)-\xi)\\
&+\frac{\pi T_L^2}{12\varepsilon_0'(k_0)}\left((f'_q(-k_0)-q'(-k_0))+A(k_0) (f_q(-k_0)-f(-k_0)) \right) H(-v^0_\rho(k_0)-\xi)\\
&+\frac{\pi T_R^2}{12\varepsilon_0'(k_0)}\left((f'_q(k_0)-f'(k_0))-A(k_0) (f_q(k_0)-f(k_0)) \right) H(\xi-v^0_\rho (k_0))\\
&+\frac{\pi T_R^2}{12\varepsilon_0'(k_0)}\left((f'_q(-k_0)-f'(-k_0))+A(k_0) (f_q(-k_0)-f(-k_0)) \right) H(\xi+v^0_\rho (k_0)).\\
\end{aligned}
\end{equation}
Finally, we have the low-temperature expansion for currents
\begin{equation}
\begin{aligned}
&j(\xi)=\frac{\pi }{12\varphi_0'(\Lambda_0)}\left(T_L^2g'_q(\Lambda_0)+T_R^2g'_q(-\Lambda_0)-(T_L^2g_q(\Lambda_0)-T_R^2g_q(-\Lambda_0))B(\Lambda_0)\right) H(v^0_\sigma (\Lambda_0)-|\xi|)\\
+&\frac{\pi }{12\varphi_0'(\Lambda_0)}\left(T_L^2g'_q(\Lambda_0)+T_L^2g'_q(-\Lambda_0)-(T_L^2g_q(\Lambda_0)-T_L^2g_q(-\Lambda_0))B(\Lambda_0)\right) H(-v^0_\sigma (\Lambda_0)-\xi)\\
+&\frac{\pi }{12\varphi_0'(\Lambda_0)}\left(T_R^2g'_q(\Lambda_0)+T_R^2g'_q(-\Lambda_0)-(T_R^2g_q(\Lambda_0)-T_R^2g_q(-\Lambda_0))B(\Lambda_0)\right) H(\xi-v^0_\sigma (\Lambda_0))\\
+&\frac{\pi }{12\varepsilon_0'(k_0)}\left(T_L^2f'_q(k_0)+T_R^2f'_q(-k_0)-(T_L^2f_q(k_0)-T_R^2f_q(-k_0))A(k_0) \right) H(v^0_\rho (k_0)-|\xi|)\\
+&\frac{\pi }{12\varepsilon_0'(k_0)}\left(T_L^2f'_q(k_0)+T_L^2f'_q(-k_0)-(T_L^2f_q(k_0)-T_L^2f_q(-k_0))A(k_0) \right) H(-v^0_\rho(k_0)-\xi)\\
+&\frac{\pi }{12\varepsilon_0'(k_0)}\left(T_R^2f'_q(k_0)+T_R^2f'_q(-k_0)-(T_R^2f_q(k_0)-T_R^2f_q(-k_0))A(k_0) \right) H(\xi-v^0_\rho (k_0)).
\end{aligned}
\end{equation}
For energy current
\begin{equation}
j_e(\xi)=\frac{\pi }{12}\left(T_L^2-T_R^2\right) H(v^0_\sigma (\Lambda_0)-|\xi|)+\frac{\pi }{12}\left(T_L^2-T_R^2\right) H(v^0_\rho (k_0)-|\xi|).
\end{equation}

We now consider the region  $\xi\pm v^0_\rho(k_0)\sim O(T)$ and $\xi\pm v^0_\sigma(\Lambda_0)\sim O(T)$, where we set $T_L=T,T_R=rT$, and we need to be careful about the singularities in the Heaviside step functions in (\ref{final_solutions}) at zero temperature because the Sommerfeld expansion will be different.
In the region  $\xi\pm v^0_\rho(k_0)\sim O(T)$, we have 
\begin{equation}
\begin{aligned}
&\int_{-\infty}^{\infty}dkn_\rho(k,\xi)p(k)=\int_{-k_0}^{k_0}p(k)dk+\\
&\frac{sign(v^0_\rho{'}(k_0))T}{\varepsilon_0'(k_0)}\left( p(k_0)D_r\left(\varepsilon_0'(k_0)\frac{\xi-v^0_\rho(k_0)}{Tv^0_\rho{'}(k_0)} \right)-p(-k_0)D_r\left(\varepsilon_0'(k_0)\frac{\xi+v^0_\rho(k_0)}{Tv^0_\rho{'}(k_0)} \right)\right) 
\end{aligned}
\end{equation}
where
\begin{equation}
D_{\eta}\left(z\right)\equiv\ln\left(1+e^{z}\right)-\eta\ln\left(1+e^{z/\eta}\right).
\end{equation}
We have a similar result  for $\int_{-\infty}^{\infty}d\Lambda n_\sigma(\Lambda,\xi)q(\Lambda)$ in the region $\xi\pm v^0_\sigma(\Lambda_0)\sim O(T)$. The whole process of low-temperature expansion is the same as in the region $\lim\limits_{T\to0}|\xi\pm v^0_\rho(k_0)|\ne0$ and $\lim\limits_{T\to0}|\xi\pm v^0_\sigma(\Lambda_0)|\ne0$. Here, we just  present the final results. For  the region $\xi\pm v^0_\rho(k_0)\sim O(T)$
\begin{equation}
\begin{aligned}
q(\xi)=q_0+T\frac{ \text{sign}(v^0_\rho{'}(k_0))}{2\pi v^0_\rho(k_0)}\left( f_q(k_0)D_r\left(\varepsilon_0'(k_0)\frac{\xi-v^0_\rho(k_0)}{Tv^0_\rho{'}(k_0)}\right)-f_q(-k_0)D_r\left(\varepsilon_0'(k_0)\frac{\xi+v^0_\rho(k_0)}{Tv^0_\rho{'}(k_0)} \right)\right)  +O(T^2),
\end{aligned}
\end{equation}
\begin{equation}
\begin{aligned}
j(\xi)=T\frac{ \text{sign}(v^0_\rho{'}(k_0))}{2\pi }\left( f_q(k_0)D_r\left(\varepsilon_0'(k_0)\frac{\xi-v^0_\rho(k_0)}{Tv^0_\rho{'}(k_0)}\right)+f_q(-k_0)D_r\left(\varepsilon_0'(k_0)\frac{\xi+v^0_\rho(k_0)}{Tv^0_\rho{'}(k_0)} \right)\right) +O(T^2). 
\end{aligned}
\end{equation}

For  the region $\xi\pm v^0_\sigma(\Lambda_0)\sim O(T)$
\begin{equation}
\begin{aligned}
q(\xi)=q_0+T\frac{ \text{sign}(v^0_\sigma{'}(\Lambda_0))}{2\pi v^0_\sigma(\Lambda_0)}\left( g_q( \Lambda_0)D_r\left(\varphi_0'(\Lambda_0)\frac{\xi-v^0_\sigma(\Lambda_0)}{Tv^0_\sigma{'}(\Lambda_0)} \right)-g_q( -\Lambda_0)D_r\left(\varphi_0'(\Lambda_0)\frac{\xi+v^0_\sigma(\Lambda_0)}{Tv^0_\sigma{'}(\Lambda_0)} \right)\right) +O(T^2),
\end{aligned}
\end{equation}
\begin{equation}
\begin{aligned}
j(\xi)=T\frac{ \text{sign}(v^0_\sigma{'}(\Lambda_0))}{2\pi}\left( g_q( \Lambda_0)D_r\left(\varphi_0'(\Lambda_0)\frac{\xi-v^0_\sigma(\Lambda_0)}{Tv^0_\sigma{'}(\Lambda_0)} \right)+g_q( -\Lambda_0)D_r\left(\varphi_0'(\Lambda_0)\frac{\xi+v^0_\sigma(\Lambda_0)}{Tv^0_\sigma{'}(\Lambda_0)} \right)\right)+O(T^2). 
\end{aligned}
\end{equation}

\end{widetext}

\end{document}